\documentclass[10pt,journal,compsoc]{IEEEtran}
\usepackage{xr}
\usepackage{colortbl}
\usepackage{rotating}
\usepackage{subfig}
\usepackage{graphicx}
\usepackage{fix2col}
\usepackage{fancyvrb}
\usepackage{graphicx}
\usepackage{relsize}
\usepackage{multirow}
\usepackage{color}
\usepackage{enumitem}
\usepackage{upgreek}
\usepackage{amsmath}
\usepackage{amssymb}
\usepackage{mathtools}
\usepackage{textcomp}
\usepackage{microtype}
\usepackage{todonotes}
\usepackage{scrextend}

\usepackage{mdframed}
\usepackage{verbatim}
\usepackage{float}
\usepackage[utf8, latin1]{inputenc}
\usepackage[T1]{fontenc}
\usepackage{times}
\usepackage{url}
\usepackage{stringstrings}
\usepackage{tabularx}
\usepackage{booktabs}
\usepackage{hyphenat}

\usepackage{array}

\usepackage{tabulary}
\usepackage{threeparttable}
\usepackage{rotating} 
\usepackage{chngpage}
\usepackage[pdftex,colorlinks=true,bookmarks=false,linkcolor=black,citecolor=black,urlcolor=black]{hyperref}
\usepackage{breakurl}

\usepackage{xspace}
\usepackage[normalem]{ulem}
\usepackage[numbers,sort&compress]{natbib}
\usepackage{listings}

\usepackage{balance}
\usepackage[bottom]{footmisc}
\usepackage{ragged2e}
\usepackage{textcomp}

\usepackage[most]{tcolorbox}
\definecolor{block-gray}{gray}{0.95}
\newtcolorbox{myquote}{colback=block-gray,grow to right by=-1mm,grow to left by=-1mm,
	boxrule=0pt,boxsep=0pt,breakable}

\newcommand{\insight}[1]{\begin{myquote}\textit{#1}\end{myquote}}  

\newcommand{\Figref}[1]{Figure~\ref{fig:#1}}
\newcommand{\figref}[1]{Fig.~\ref{fig:#1}}

\newcommand{\fign}[1]{\ref{fig:#1}}
\newcommand{\Figs}{Figures}

\newcommand{\chaps}[1]{Chapters}

\newcommand{\Tabref}[1]{Table~\ref{tab:#1}}
\newcommand{\tabref}[1]{Table~\ref{tab:#1}}

\newcommand{\secref}[1]{Sec.~\ref{sec:#1}}

\newcommand{\figlabel}[1]{\label{fig:#1}}
\newcommand{\tablabel}[1]{\label{tab:#1}}
\newcommand{\seclabel}[1]{\label{sec:#1}}

\newcounter{rqstcounter}

\newcounter{hppcounter}

\newcounter{patternc}
\newcommand{\hashifdef}{\texttt{\small{{\#ifdef}\xspace}}}
\newcommand{\ifndef}{\texttt{\small{{\#ifndef}\xspace}}}

\newcommand{\auroc}{AUROC\xspace}

\newcommand\parhead[1]{\vspace{.26mm}\noindent\textit{{#1.}}}  
\newcommand{\rodrigo}{Queiroz et al.\xspace}
\newcommand{\metRod}{\textsc{QueirozMet}}
\newcommand{\metA}{\textsc{ProcMet}}
\newcommand{\metB}{\textsc{ProcStructMet}}

\newcommand{\foperands}[1]{\texttt{\small{#1}\xspace}}
\newcommand{\mtnote}[1]{\textsuperscript{\TPTtagStyle{#1}}}
\newcommand\theader[1]{\textsf{{#1}}}

\newcommand\metricName[1]{\texttt{\textsc{#1}}}

\newcommand{\feat}{\textit{feat}}

\newboolean{showcomments}
\setboolean{showcomments}{true} 
\ifthenelse{\boolean{showcomments}}
{\newcommand{\nb}[2]{
		\fcolorbox{black}{yellow}{\bfseries\sffamily\scriptsize#1}
		{\sf\small$\blacktriangleright$\textit{#2}$\blacktriangleleft$}
	}
	
}
{\newcommand{\nb}[2]{}
	
}

\usepackage[normalem]{ulem} 
\usepackage{xcolor}

\begin{document}

\title{Feature-Oriented Defect Prediction:\\Scenarios, Metrics, and Classifiers}

\author{Mukelabai~Mukelabai, Stefan~Str\"uder, Daniel~Str\"uber, Thorsten~Berger}

\IEEEtitleabstractindextext{%
\begin{abstract}
	Software defects are a major nuisance in software development and can lead to considerable financial losses or reputation damage for companies. To this end, a large number of techniques for predicting software defects, largely based on machine learning methods, has been developed over the past decades. These techniques usually rely on code-structure and process metrics to predict defects at the granularity of typical software assets, such as subsystems, components, and files. In this paper, we systematically investigate \textit{feature-oriented} defect prediction: predicting defects at the granularity of features---domain-entities that abstractly represent software functionality and often cross-cut software assets.
	Feature-oriented prediction can be beneficial, since: (i) particular features might be more error-prone than others, (ii) characteristics of features known as defective might be useful to predict other error-prone features, and (iii) feature-specific code might be especially prone to faults arising from feature interactions.
	
	We explore the feasibility and solution space for feature-oriented defect prediction. We design and investigate scenarios, metrics, and classifiers. Our study relies on 12 software projects from which we analyzed 13,685 bug-introducing and corrective commits, and systematically generated 62,868 training and test datasets to evaluate the designed classifiers, metrics, and scenarios. The datasets were generated based on the 13,685 commits, 81 releases, and 24, 532 permutations of our 12 projects depending on the scenario addressed.  We covered scenarios, such as \textit{just-in-time (JIT)} and \textit{cross-project} defect prediction. Our results confirm the feasibility of feature-oriented defect prediction. We found the best performance (i.e., precision and robustness) when using the Random Forest classifier, with process and structure metrics. Surprisingly, we found high performance for single-project JIT (median \auroc $\ge 95\,\%$) and release-level (median \auroc $\ge 90\,\%$) defect prediction---contrary to studies that assert poor performance due to insufficient training data. Lastly, we found that a model trained on release-level data from one of the twelve projects could predict defect-proneness of features in the other eleven projects with median performance of 82\,\%, without retraining on the target projects. Our results suggest potential for defect-prediction model-reuse across projects, as well as more reliable defect predictions for developers as they modify or release software features.

\end{abstract}

}

\maketitle

\IEEEraisesectionheading{\section{Introduction}\label{sec:intro}}

\looseness=-1
\IEEEPARstart{S}{oftware} errors are a significant cause of financial and reputation damage to companies. Such errors range from minor bugs to serious security vulnerabilities. In this light, it is preferable to warn developers about such problems early, for instance, when releasing updated software that may be affected by errors. 

\looseness=-1
Over the past two decades, a large variety of techniques for error detection and prediction has been developed, largely based on machine learning techniques\,\cite{Challagulla2008}. These techniques use historical data of \textit{defective} and \textit{clean} (defect-free) changes to software systems in combination with a carefully compiled set of attributes (a.k.a., \textit{features}\footnote{To avoid ambiguity, throughout this paper, we use the term ``attribute'' instead of ``feature'' to describe dataset characteristics in the context of machine learning.}) to train a given classifier\,\cite{Alsaeedi2019,Hammouri2018}. This information about the past can then be used to make an accurate prediction of whether a new change to a piece of software is defective or clean. Choosing an algorithm for classification is a nontrivial issue. Studies show that, out of the pool of available algorithms, both tree-based (e.g., J48, CART or Random Forest) and Bayesian algorithms (e.g., Na\"{\i}ve Bayes (NB), Bernoulli-NB or multinomial NB) are the most widely used\,\cite{Son2019}. Alternatives include logistic regression, k-nearest-neighbors or artificial neural networks\,\cite{Challagulla2008}. 
The vast majority of existing work uses these techniques for defect prediction at the granularity of sub-systems, components, and files, and does not come to a definitive consensus on their usefulness---the ``best'' classifier generally seems to depend on the prediction setting.

\looseness=-1
We present a systematic investigation of \textit{feature-oriented defect prediction}, a term that we define as defect prediction at the  granularity of features (i.e., predicting error-prone features---\textit{feature-level} defect prediction) and/or using feature information to predict defects, potentially at other granularity levels than features (\textit{feature-based} defect prediction).
Features are a primary unit of abstraction in software product lines and configurable systems\,\cite{Apel2013,berger2015feature,kang.ea:1990:foda,damir2019principles}, but also play a crucial role in agile development processes, where organizations strive towards feature teams and organize sprints around feature requests, for shorter release cycles\,\cite{larman2008scaling}. Notably, features abstract over traditional software assets (e.g., source files) and often cross-cut them\,\cite{passos.ea:2018:tse}, constituting more coherent entities from a domain perspective.

Feature-oriented\,defect\,prediction is promising for several reasons:
First, since a given feature  might be historically more or less error-prone, a change that updates  the feature may be more or less error-prone as well.
Second, features that are more or  less likely to be error-prone might have certain characteristics that can be harnessed to predict defects.
Third, feature-specific code might be especially prone to faults arising from feature interactions\,\cite{bruns2005foundations,zave:2004:features,apel2014feature}.

\looseness=-1
We address typical as well as new scenarios for defect prediction.
While \textit{feature-level} defect prediction is a new scenario, we also investigate if  feature information can lead to enhancements in the traditional scenario of \textit{file-level} defect prediction.
To address two typical granularity levels of changes, we consider both \textit{release-level} and \textit{commit-level (a.k.a., just-in-time)} defect prediction.
Beyond the \textit{within-project} scenario, where training and prediction are performed in the same project, we also address \textit{cross-project} defect prediction, in which models are reused over several different software projects.

Our research questions seek to investigate (i) metrics and classifiers that are best suited for feature-level defect prediction, and (ii) whether feature-based metrics and classifiers can lead to improvements in the traditional scenarios of file-level, commit-level, release-level, and cross-project defect prediction. 
Towards the former (RQ1), we investigate the design space of metrics and classifiers for feature-level defect prediction:

\smallskip
\noindent RQ1: \textit{What combination of metrics and classification algorithms yields the best performance for feature-level defect prediction?} We analyzed a total of 14 metrics (divided into three subsets) and 7 classifiers to understand what effect different classifiers and metrics have on prediction quality.

\smallskip
The results of RQ1 formed the basis for the analysis we performed in the subsequent research questions RQ2--5, which were devoted to traditional defect prediction scenarios described in the literature. First, we sought to understand what impact feature-based metrics may have on file-level defect prediction:

\smallskip
\noindent RQ2: \textit{What is the effect of feature-based metrics on file-level defect prediction?} We generated a file-level defect prediction dataset from our subject systems by using 17 file-level metrics proposed in Moser et al.'s widely impactful paper \,\cite{Moser2008}. 
We then analyzed the classifier performance resulting from running predictions on the file-based dataset with feature metrics and without feature metrics.

\smallskip
Next, we compared the performance of feature-based and file-based defect prediction:

\smallskip
\noindent RQ3: \textit{How does feature-based defect prediction perform compared to file-based defect prediction?} 
We compared the proportion of defective files correctly predicted by either method.

\smallskip
Next, we sought to understand the effectiveness of feature-based prediction when performed for individual projects at each commit or release. This is especially important considering the potential for insufficient training data, as commonly reported\,\cite{kamei2016studying,zimmermann2009cross}:

\smallskip
\noindent RQ4: \textit{To what extent can feature-based defect prediction support developers as they modify or release software features?} We analyzed the performance of commit- and release-level predictions of each project, by generating training data from all previous commits/releases of a project to predict defect proneness of features in each subsequent commit/release of the same project.

\smallskip
Finally, we considered cross-project defect prediction:

\smallskip
\noindent RQ5: \textit{To what extent can feature-based defect prediction models be reused across projects without re-training?} 
We analyzed results from a total of 24, 532 combinations of our 12 projects that provided training data to predict the defect-proneness of features in each of the other projects not in the training set. For instance, we used training data from 1 project to predict for each of the remaining 11, or we used training data from 4 projects to predict for each of the remaining 8 projects, and so on.
\smallskip

\looseness=-1
Our results show the feasibility and benefits of feature-oriented defect prediction.
For RQ1 (metrics and classifiers), we observed higher performance (above 85\,\%) when using the Random Forest classifier in combination with both process and structure metrics than when using process metrics only. With respect to RQ2 (impact of metrics on file-level prediction), we did not find a statistically significant performance benefit resulting from adding feature metrics to file-level datasets. However, using an attribute ranking algorithm, we found that all 14 feature-based metrics were ranked in the top 75\,\% (i.e., 24 of 31) best metrics for the file-level datasets and that of the top 10 (30\,\%) metrics, half of them were feature-based. For RQ3 (feature- vs. file-based defect prediction), we obtained higher performance values with feature-based defect prediction than with file-based. Furthermore, while the former correctly predicted 274 (65\,\%) out of 419 defective features mapped to 628 defective files, the latter failed to correctly predict a single one of those defective files. With respect to RQ4 (single-project change-based defect prediction), we observed very high performance values of over 95\,\% for commit-level predictions and over 90\,\% for release-level predictions on individual projects, which promises effective defect prediction that is in-line with software development activities. Lastly, for RQ5 (cross-project prediction), we found that feature-based prediction models can be reused on different projects without retraining, with better performance obtained for release-level datasets than commit-level datasets. Even though the best performance was obtained when using training data from a combination of 11 of the 12 projects, we found that in some cases, a single project was sufficient to train on; with a median performance of over 75\,\% for all combinations of training on a single project and evaluating on the remaining 11 projects.

In summary, we contribute:
\begin{itemize}[leftmargin=*]
	\item A \textit{dataset for feature-based defect prediction.}
	The dataset is based on 12 projects and contains features in specific versions, labeled as either \textit{defective} or \textit{clean}.
	Feature information was extracted from preprocessor macros (\hashifdef~and~\ifndef) in the projects' source code files.
	The labels were determined using existing automated heuristics targeting file-based defect prediction, which we refined to obtain more accurate results in the considered projects. 
	
	\item A \textit{set of 14 feature-based metrics} designed for training machine learning classifiers to predict software defects at the granularity of features. The set consists of 8 process and 6 structure metrics.
	
	\item An \textit{evaluation of feature-based defect prediction} with respect to its efficacy over file-based predictions, its ability to predict defect proneness of features as developers modify or release product features, and its potential for reuse across different projects without retraining classification models.
	
	\item A \textit{replication package} with all data and code, publicly available in our online appendix\,\cite{appendix:Online}.	
\end{itemize}

This manuscript significantly extends a previous conference paper\,\cite{struder2020feature} that focused on analyzing metrics and classifiers for feature-based defect prediction (RQ1). With our extension, we broaden the scope of our work by performing a comprehensive study of new and existing feature-based defect prediction scenarios (RQ2--RQ5). We investigate whether using feature-based metrics on the traditional file-based dataset improves the predictions (RQ2), how well feature-based prediction models perform compared to the traditional file-based ones (RQ3), how effective feature-based predictions would be when performed as developers change (during commits) or release software features (RQ4), and to what extent our novel prediction models can be reused across projects (RQ5).

\begin{figure*}[ht!]
	\centering
	\includegraphics[width=\textwidth]{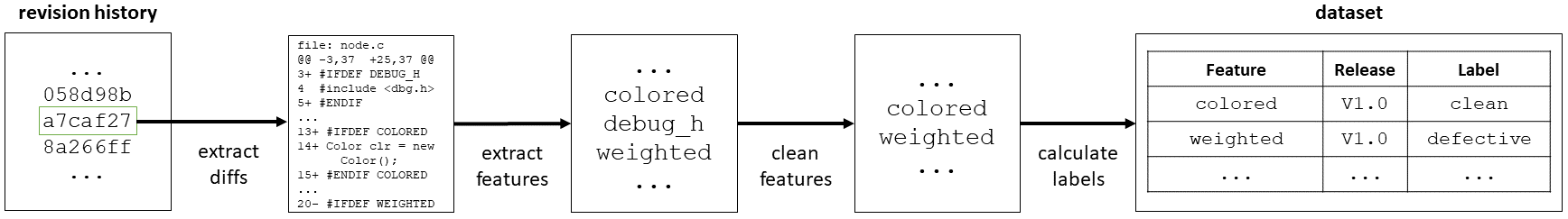}
	\vspace{-.7cm}
	\caption{Dataset creation\figlabel{fig:dataset}}
	\vspace{-.3cm}
\end{figure*}

\section{Background and Related Work}
\seclabel{sec:relwork}

\looseness=-1
\parhead{Software defect prediction} Defect prediction is an active research area in software engineering that has been studied for the past five decades\,\cite{nam2013transfer,rahman2011bugcache,d2010extensive,zimmermann2008predicting,menzies2006data}. The earliest studies date back to the 1970s and were done by Akiyama\,\cite{akiyama1971example}, McCabe\,\cite{mccabe1976complexity}, and Halstead\,\cite{halstead1977elements}, who used code complexity metrics to estimate defects (without machine learning). The vast majority of recent studies relies on machine learning techniques\,\cite{d2012evaluating,bird2011don,lee2011micro,bacchelli2010popular,hassan2009predicting,menzies2006data,nagappan2005use} and follows standard procedures of (i) extracting instances (dataset records) from software archives based on the chosen granularity level (e.g., file,  class, or method level), (ii) labeling the instances (e.g., as defective or clean) and applying metrics, (iii) optionally applying preprocessing techniques, such as feature selection\,\cite{shivaji2012reducing} or normalization\,\cite{menzies2006data}, and (iv) making predictions for unknown instances---predicting either bug-proneness of source code (classification) or the number of defects in source
code (regression).

\looseness=-1
To characterize the defect-proneness of source code, several metrics have been proposed, including structure and process metrics. 
While structure metrics generally measure the complexity and size of code, process metrics quantify several aspects of the development process, such as changes of source code, code ownership, and developer interactions. 
The onset of version control systems has facilitated the application of process metrics to defect prediction\,\cite{Rahman2013,lee2011micro,hassan2009predicting,Moser2008}, which have been demonstrated to outperform structure metrics in many cases\,\cite{Rahman2013,Moser2008,nagappan2005use}. Different measures are used to assess the performance of classification models, the most common being precision, recall, and F-score (see \secref{subsec:metrics}). However, since most prediction models predict probabilities of defect-proneness, these measures require the use of a minimum probability threshold to declare an instance defective or not. Such performance measures that require the use of threshold values are discouraged\,\cite{lessmann2008benchmarking},  since results may vary and are hard to reproduce\,\cite{mende2010replication}. 
A more reliable, threshold-invariant metric is the area under the receiver operating characteristic curve (AUC-ROC). It plots the true positive rate against the false positive rate taking into account all possible threshold values (between 0 and 1). Thus, AUC-ROC indicates how much a prediction model is capable of distinguishing between classes. Furthermore, the area under cost-effective curve (AUCEC)\,\cite{rahman2011bugcache,arisholm2007data} is sometimes used to measure how many defects can be found in the top \textit{n\%} lines of code so as to provide priorities to quality assurance teams and developers.

\looseness=-1
Defect prediction models may target quality assurance before product release\,\cite{Rahman2013} (a.k.a., \textit{release-based}), or prediction of defects whenever the source code is changed (i.e., predicting bug-inducing changes, a.k.a., \textit{just-in-time (JIT)} defect prediction models\,\cite{kamei2016studying,kamei2012large,kim2008classifying}). In general, JIT models suffer from insufficient training data. To overcome this limitation for new projects or projects with less historical data, the notion of cross-project defect prediction has been studied as well\,\cite{nam2013transfer,he2012investigation,zimmermann2009cross}. To achieve better performance, cross-project predictions generally require careful selection of training data\,\cite{kamei2016studying,zimmermann2009cross}, e.g., from similar projects to create a large training dataset, or they require ensembles of models from several projects.

\looseness=-1
Defect prediction models have been constructed at various granularity levels, including sub-system\,\cite{klas2010transparent,hassan2009predicting,fenton2008effectiveness}, component/package\,\cite{zimmermann2008predicting,lessmann2008benchmarking,nagappan2005use}, file/class\,\cite{nam2013transfer,zimmermann2009cross,marcus2008using,menzies2006data,ostrand2005predicting}, method\,\cite{hata2012bug,giger2012method}, and change (hunk)\,\cite{kim2008classifying} level. Only one study\,\cite{Queiroz2016} has considered the feature granularity level, in a minimalistic setup with one software project, 5 process metrics, and 3 classifiers. Our study investigates defect prediction in a much comprehensive setting, considering several defect prediction scenarios, 14 feature-based process and structure metrics to characterize the defect-proneness of features. We use data from 12 pre-processor-based projects and 7 classifiers. We rely on the AUC-ROC (also simply called \auroc) measure to assess the performance of our selected classifiers.

\looseness=-1
\parhead{Machine learning and software product lines} 
Defect prediction presents a natural application avenue for machine learning in product line engineering (a paradigm for engineering variant-rich software systems\,\cite{apel:2013:fospl}).
Most existing work in this area has focused on the sampling of configurations for various use cases;
the recent survey by Pereira et al.\,\cite{pereira2019learning} provides an overview.
Focusing on performance predictions, Siegmund et al.\,\cite{siegmund:splc:2011} use machine learning and sampling techniques to build performance influence models, quantifying the performance impact of specific features and interactions.
Temple et al.\,\cite{Temple2016} use machine learning to infer missing product line constraints\,\cite{nadi.ea:2015:tse}, based on a random sampling of products and an oracle that assesses whether a particular configuration leads to a valid product. 
In contrast to us, they are interested in faulty feature combinations, rather than erroneous features.
The same authors also investigated the use of learned adversarial configurations in the context of quality assurance\,\cite{temple2019towards}.
Considering the reverse direction of applying variability concepts to machine learning, Ghofrani et al.\,\cite{ghofrani2019applying} propose to investigate product lines of deep neural networks, which establish reuse of existing trained networks by identifying features and composing them.
They investigate the reuse potential in an associated empirical study\,\cite{ghofrani2019reusability}.

\parhead{Feature metrics} Several variability-aware feature metrics have been proposed in the literature\,\cite{el2019metrics,passos.ea:2013:evolution,Liebig2010,Berger:2014:TSA:2556624.2556641} that measure characteristics of feature specifications (variability models)\,\cite{kang.ea:1990:foda,damir2019principles}, code, or of the mapping between the feature specification and code artifacts. These metrics target specific variability implementation mechanisms, typically classified\,\cite{strueber2020modelsvar} into annotative mechanisms---such as the C preprocessor (e.g., \hashifdef\,\cite{Medeiros:2015wg})---and compositional mechanisms---such as feature modules (e.g., AHEAD\,\cite{Batory:2004bw}).
Using metrics allows conceiving lightweight analysis techniques\,\cite{mukelabai.ea:2018:analysis} for complex systems, such as product lines.

Common among annotation-based metrics are code-related metrics\,\cite{Liebig2010,hunsen2016preprocessor,Queiroz2015} that measure the nesting depth of features, lines of feature code, feature scattering degree (to what extent a feature's implementation is spread across the codebase\,\cite{passos.ea:2018:tse}), and tangling degree (to what extent a feature's implementation is mixed with that of other features, implying feature interactions\,\cite{zave:2004:features,apel2014feature,bruns2005foundations}). We use these four kinds of structural metrics to characterize the defect-proneness of a feature, hypothesizing that the higher the value of each metric is, the more likely a feature is to be defective.
All of these existing metrics are structure-based. 
With the exception of \rodrigo's defect prediction work\,\cite{Queiroz2016} (which we consider during metric engineering), no existing work proposes dedicated process metrics for features.

\section{Study Design}
\looseness=-1
To explore the design space of scenarios, metrics, and classifiers, our 
methodology comprised:
(i) creating a dataset of feature labels over the history of 12 software projects;
(ii) creating two new metric sets designed for feature-oriented defect prediction;
 (iii) selecting and training 7 classifiers we considered for our evaluation; and (iv) evaluating 5 different defect prediction scenarios, such as just-in-time and cross-project defect prediction.

\looseness=-1
To navigate this design space without considering all combinations of these three dimensions, we chose the following strategy: 
First, for a fixed scenario, we evaluated all combinations of metrics and classifiers. We identified a consistently high-performing combination of metrics and classifier. Thereafter, we used this combination of metrics and classifier to thoroughly evaluate the additional scenarios considered in this work.

\subsection{Dataset Creation}
\seclabel{sec:dataset}

To create a machine learning model for predicting the defect-proneness of software features, we created training and test datasets whose instances are features. This is in contrast with commonly used granularity levels such as components, files or methods. To this end, we relied on software projects with available revision histories and that use preprocessor macros (e.g., \hashifdef) to annotate source code with features. We followed the process outlined in \figref{fig:dataset} to extract feature references (by pattern matching preprocessor macros) in files that were modified during commits. We labeled these features as defective or clean based on whether one or more files implementing each feature were identified to be defective. Below, we describe this process in more detail.

\parhead{Software projects} 
We generated datasets based on data from the full revision histories of 12 preprocessor-based software projects---these projects have been subjects of prior research on features and software product lines\,\cite{Hunsen2015,Liebig2010,Queiroz2015,Queiroz2016}. From these four publications, we obtained an initial set of 44 projects, which we filtered by applying the following inclusion criteria: First, the project's source code uses preprocessor directives as variability mechanism. Second, meta-data on release versions is available in the form of several tags specifying release versions.
Third, the project has a nontrivial (greater than 5) number of features.
Fourth, the project's commit messages are given in English---a prerequisite for the heuristics we used for detecting bug-fixing commits. We checked these criteria manually, yielding a selection of 12 projects, which we list in \tabref{tab:datasetexample}, together with context, repository sources, and additional information.

\setlength{\tymin}{3.5cm}
\begin{table*}[tb]
	\fontsize{7}{7}\selectfont
	\centering
	\caption{Subject systems}
	\vspace{-.3cm}
	\tablabel{tab:datasetexample}
	\begin{threeparttable}
		\fontfamily{ptm}\selectfont
		
		\begin{tabulary}{\textwidth}{@{}LLRRRLLCL@{}} 
			
			\toprule
			\theader{project}	&	\theader{description}	&	\theader{corrective commits}	&	\theader{bug-introducing commits}	
			&	\theader{features}	&	\theader{training-set releases}	
			&	\theader{test-set releases}	&	\theader{split ratio\tnote{1}	}	&	\theader{URL}\\
			\midrule
			\textbf{Blender}  & 3D-modeling tool &7,760&3,776&1,400&2.70 - 2.77                                                         & 2.78 - 2.80                                                     & $73:27$& \url{github.com/sobotka/blender}          \\
			\textbf{Busybox}  & UNIX toolkit     &1,236&802&628&1\_16\_0 - 1\_25\_0                                                 & 1\_26\_0 - 1\_30\_0                                             & $71:29$  & \url{git.busybox.net/busybox/}          \\
			\textbf{Emacs}    & text editor      &4,269&2,532&718&25.0 - 26.0                                                         & 26.1 - 26.2                                                     & $71:29$ & \url{github.com/emacs-mirror/emacs}          \\
			\textbf{GIMP}     & graphics editor  &1,380&854&204&2\_8\_2 - 2\_10\_4                                                  & 2\_10\_6 - 2\_10\_12                                            & $71:29$   & \url{gitlab.gnome.org/GNOME/gimp}          \\
			\textbf{Gnumeric} & spreadsheet      &1,498&1,191&637&1\_10\_0 - 1\_12\_10                                                & 1\_12\_20 - 1\_12\_30                                           & $75:25$  & \url{gitlab.gnome.org/GNOME/gnumeric}          \\
			\textbf{gnuplot}  & plotting tool    &854&1,215&558&4.0.0 - 4.6.0                                                       & 5.0.0                                                           & $80:20$     & \url{github.com/gnuplot/gnuplot}          \\
			\textbf{Irssi}    & IRC client       &52&22&9&1.0.0 - 1.0.4                                                       & 1.0.5 - 1.0.6                                                   & $71:29$    & \url{github.com/irssi/irssi}          \\
			\textbf{libxml2}  & XML parser       &324&88&200&2.9.0 - 2.9.7                                                       & 2.9.8 - 2.9.9                                                   & $80:20$   & \url{gitlab.gnome.org/GNOME/libxml2}          \\
			\textbf{lighttpd} & web server       &1,078&929&230&1.3.10 - 1.4.20                                                     & 1.4.30 - 1.4.40                                                 & $67:33$  & \url{git.lighttpd.net/lighttpd/lighttpd1.4.git/}          \\
			\textbf{MPSolve}  & polynom solver   &151&211&54&3.0.1 - 3.1.5                                                       & 3.1.6 - 3.1.7                                                   & $75:25$ & \url{github.com/robol/MPSolve}         \\
			\textbf{Parrot}   & virtual machine  &3,109&3,072&397&1\_0\_0 - 5\_0\_0                                                   & 6\_0\_0 - 7\_0\_0                                               & $71:29$    & \url{github.com/parrot/parrot}         \\
			\textbf{Vim}      & text editor      &371&696&1,158&7.0 - 7.4                                                           & 8.0 - 8.1                                                       & $71:29$      & \url{github.com/vim/vim}  \\
			
			\bottomrule				
		\end{tabulary}
		\begin{tablenotes}[para]
		
			~\item[1] percentage of training and test releases		
		\end{tablenotes}
	\end{threeparttable}
	\vspace{-0.3cm}
\end{table*}

\parhead{Retrieval} 
To retrieve our subject projects' revision histories, we used the library PyDriller
\,\cite{Spadini2018}). It allows easy data extraction from Git repositories to obtain commits, commit messages, commit authors, diffs, and more (called "metadata" in the following). To this end, we created Python scripts for receiving the commit metadata, including the release number to which each commit belonged.

\looseness=-1
For each modified file within a commit, we collected metadata, such as \textit{commit hash, commit author, commit message, filename, and diff (changeset)}, that we used for calculating metrics (\secref{sec:metrics}) and labeling of instances in our datasets. This metadata was saved in a MySQL database, available as part of our online appendix\,\cite{appendix:Online}. For each of our subject projects, we create a separate table in the database in which we store the above metadata for each file, including the name of the project and the release number associated with the commit in which the file was changed.

\begin{table}[b]
	\vspace{-.3cm}
	\centering
	\caption{Key characteristics of the dataset}
	\tablabel{tab:dataset-numbers}
	\vspace{-.3cm}
	\begin{tabular}{lrr} 
		\toprule
		instances          & defective&clean  \\ 
		\midrule
		
			$13,177$     & $2,168$   & $11,009$                                                           \\
	
		\bottomrule
	\end{tabular}
\end{table}

\parhead{Feature reference extraction and cleaning} 
Using regular expressions, we extracted feature references in each modified file within a commit changeset, by pattern-matching the preprocessor macros \hashifdef~and\,\ifndef. Combinations of features (e.g., \hashifdef\,\foperands{A \& B}) are stored in their identified form.

\looseness=-1
This way of identification has some obstacles. In some C programming paradigms, it is common to include header files in the source code using preprocessor directives, in the same way as features. However, we ignored these ``header macros,'' as they do not represent actual features.
In general, these header macros are identifiable through their suffix \texttt{\_h\_} to the name, such as \texttt{macroname\_h\_}. Through a manual review of the identified feature references, we also ignored feature references when the preprocessor directives occurred in comments.

\parhead{Label calculation} 
For each identified feature in each revision, we calculated a label, specifying if the feature is \textit{defective} or \textit{clean}. To this end, we relied on a common automated heuristic for identifying corrective and bug-introducing commits\,\cite{Zimmermann2007}. We modified it for our purpose and mapped the results to features, as explained below.
The heuristic scans commit messages for the presence of the keywords ``bug,'' ``bugs,'' ``bugfix,'' ``error,'' ``fail,'' ``fix,'' ``fixed,'' and ``fixes.'' 

In a manual inspection of the results, we noticed many false positives.
Especially in lengthy commit messages, we noticed an increased probability that our keywords are used in an irrelevant context, e.g., handling of ``fixed fonts'' in the implementation of \textit{emacs}.
We modified the heuristic to only consider the first line of each commit since the main purpose of the commit is usually stated in the first line or sentence. We then took a sample of about 50--100 commits per project (approx 500 in total) to evaluate the modified heuristic, and found that it decreased the number of false positives significantly. However, as a general limitation of our technique (similar to other techniques used in defect prediction studies) we do not guarantee that a commit does not have bugs, but instead focus on confirmed bugs specified by developers.

\looseness=-1
We used the corrective commits to identify the corresponding bug-introducing commits.
The state-of-the-art algorithm, SZZ\,\cite{Sliwerski2005,Spadini2018}, uses heuristics to identify the commits in which the lines leading to the later-fixed bug have been introduced. We used the available SZZ  implementation of PyDriller.
\Tabref{tab:datasetexample} gives an overview of the number of corrective and bug-introducing commits and the number of features identified per project.

\looseness=-1
Finally, for labeling, we first compute labels for files, and then use these labels to calculate the labels for associated features.
A file is labeled as \textit{defective} in a particular release if there is at least one bug-introducing commit that changes the file, and as \textit{clean} otherwise.
A feature is labeled as \textit{defective} in a particular release if it is associated with at least one defective file, and as \textit{clean} otherwise.
Corrective commits are not reflected directly in labels, since we are interested in the error-proneness of particular features.
Key figures giving an overview of the created dataset are listed in \tabref{tab:dataset-numbers}.

\autoref{fig:bug-example} shows the diffs of a corrective (A) and a bug-introducing (B) commit to a feature \texttt{FEAT\_TEXT\_PROP} from the project \textit{Vim}. The diff of commit A shows that the arguments of the method call \texttt{vim\_memset} have been replaced. According to the associated commit message, the original method call caused a "memory access error." Commit A was, therefore, identified as corrective because the commit message contains the keyword "error." To identify the bug-introducing commit B of the file concerned, we specify the hash of the corrective commit A to the SZZ algorithm. In its portion of the diff, we can see that commit B has put the feature \texttt{FEAT\_TEXT\_PROP} in the file with the incorrect method call. Consequently, we consider the commit to be bug-introducing, and the associated file and feature to be defective in that particular release.

\subsection{Evaluation Metrics}
\seclabel{subsec:metrics}
To compare the classifiers with regard to prediction quality, we consider two types of evaluation metrics, commonly used for this purpose in the field of information retrieval\,\cite{Sammut2017}.
First, \textit{precision}, \textit{recall}, and \textit{F-score}, which quantify information about the percentage of true and false predictions, based on an available \textsl{confusion matrix}.
Second, \textit{receiver operating characteristic} (ROC) curves and the associated \textit{area under curve} (AUC), which provide a visual and more robust way for assessing prediction quality than confusion-matrix-based metrics.

All our evaluation metrics assume a \textit{ground truth}, specifying for each given class the entries that belong to it (positives) and those that do not (negatives).
In our case, entries are features with regard to a given release.
The classes are \textit{defective} and \textit{clean}.
The ground truth was constructed in the labeling step during dataset construction (see \secref{sec:dataset}).

\parhead{Recall, Precision, and F-score} 
We follow the standard definition of precision, recall, and F-score.
Intuitively, recall quantifies how exhaustively the classifier identified all entries of the class, comprised of true positives (TP) and false negatives (FN), respectively. 
Precision quantifies the percentage of true positives (TP) among all entries assigned to a particular class (also including false negatives, FN).
The F-score is the harmonic mean of precision and recall, representing a balance between both.
In contrast to other confusion-based-matrix (e.g., accuracy), these metrics are considered as useful on imbalanced datasets, such as ours.
These metrics are computed as follows:

\smallskip

\noindent{}$\text{Recall} = \frac{TP}{TP+FN}$, \ \ \ $\text{Precision} = \frac{TP}{TP+FP}$, \ \  \ $\text{F-score} = \frac{2TP}{2TP+FP+FN}$ 

\smallskip

\medskip
\noindent{}\textbf{ROC-AUC.} 
We determined the ROCs and AUCs of the individual classifiers. These have the benefit that they represent performance in a visual, understandable way, while at the same time making the quality assessment more robust:
Precision, recall, and F-score depend on a predefined threshold, which is used in the classifiers to assign each instance to a class.
A robust classifier shows good predictive ability regardless of the chosen threshold value.

ROC curves encode this intuition, by describing the relationship between the TP rate (a.k.a. recall, y axis) and the FP rate (x axis), indicating the proportion of predictions that are incorrectly evaluated as positive\,\cite{Sammut2017,Alpaydin2010}. 
The FP rate is calculated as follows:

\smallskip
\begin{center}
	\noindent{}\text{FP rate} = $\frac{FP}{FP+TN}$
\end{center}

\medskip
Datapoints on the curve are obtained by taking into account all possible values for the threshold that determines when an instance is assigned to a particular class. 

The AUC area indicates the extent to which a classifier is able to make correct predictions under a changing threshold value. The higher this value is, the more robust the classifier is in making correct predictions.
The ideal value is 1.0, whereas a value of 0.5 indicates a predictive ability on the same level as random guessing.

Since RQ1 focuses on evaluating metrics and classifiers, we present all evaluation metrics (precision, recall, f-score, and ROC) in our results. We find a large agreement between the different evaluation metrics in RQ1.
Consequently, in RQ2---5, due to the above mentioned benefits of \auroc, we only report \auroc.

\subsection{Evaluation of scenarios}

We systematically evaluated feature-based defect prediction in different scenarios, corresponding to our five research questions.

In RQ1, we considered the scenario of \textit{feature-level defect prediction}. Following our strategy of evaluating combinations of metric sets and classifiers in this fixed scenario, we analyzed the performance of seven classifiers in combination with three sets of metrics: the first set (\metRod) from the literature\,\cite{Queiroz2016} comprising five process metrics; the second (\metA), which we engineered, comprising eight process metrics; and the third (\metB), which we also engineered, comprising  the eight process metrics from \metA\, plus six structural metrics. We analyzed the effect of the three sets and individual metrics on prediction quality, as well as identified the best performing classifier that we used to address RQ2---5.

To address RQ2 (impact of feature-based metrics on file-based defect prediction), we calculated 17 file-based (\tabref{tab:fileMetrics}) and 14 feature-based metrics (\tabref{tab:featureMetrics}) for each file in our training and test datasets. We then analyzed the performance of our chosen classifier when predicting defects with and without feature-based metrics.

To address RQ3 (feature- vs. file-based defect prediction), we mapped features in our feature-based test dataset (created in RQ1) to files in the file-based test dataset (created in RQ2) and then analyzed the proportion of defective files correctly predicted by feature-based predictions and by file-based predictions. Both the feature-and file-based datasets were created from the same set of releases of our subject projects (see \tabref{tab:datasetexample}).

In RQ4 (change- and release-based defect prediction) we generated datasets at commit- and release-level for individual projects (unlike previously done in RQ1 and RQ2) and made predictions per commit or release. Since our aim in RQ1 was to identify the best set of metrics and the best classifier, we combined data from all projects and releases and split the data into training and test sets as shown in (see \tabref{tab:datasetexample}). In RQ4, we aimed at understanding to what extent feature-based prediction can support developers as they modify or release features in their individual projects. Hence, we generated training data from all commits or releases up to the \textit{n}th, and predicted the defect proneness of features changed in the subsequent (\textit{n+1}th) commit or release.

In RQ5 (cross-project defect prediction) we generated training data from different combinations of projects and created a model of our classifier that we used to predict defects in projects in the test set. Unlike RQ1, here we make predictions for projects whose data has not been exposed to the classifier.

\section{Metric Sets and Classifiers for Feature-Level Defect Prediction (RQ1)}

We now address RQ1 (\textit{What combination of metrics and classification algorithm yields best performance for feature-level defect prediction?}). We present how we addressed the research question and later discuss our results and their implications.
\looseness=-1
\subsection{Methodology}
\seclabel{sec:rq1method}
We first describe how we engineered attributes of our feature-based defect datasets, followed by how we selected and trained classifiers.
\subsubsection{Selection of Metrics}
\seclabel{sec:metrics}

Selecting an effective set of attributes for classifier training (a.k.a., \textit{feature engineering}) is commonly considered the decisive factor for the success or failure of machine learning applications\,\cite{domingos2012few}.
To reflect this crucial role, we iteratively designed a suitable set of attributes, following a design-science approach\,\cite{hevner2004design}.

\begin{figure}[b]
	\vspace{-0.3cm}
	\centering
	\includegraphics[width=0.7\linewidth]{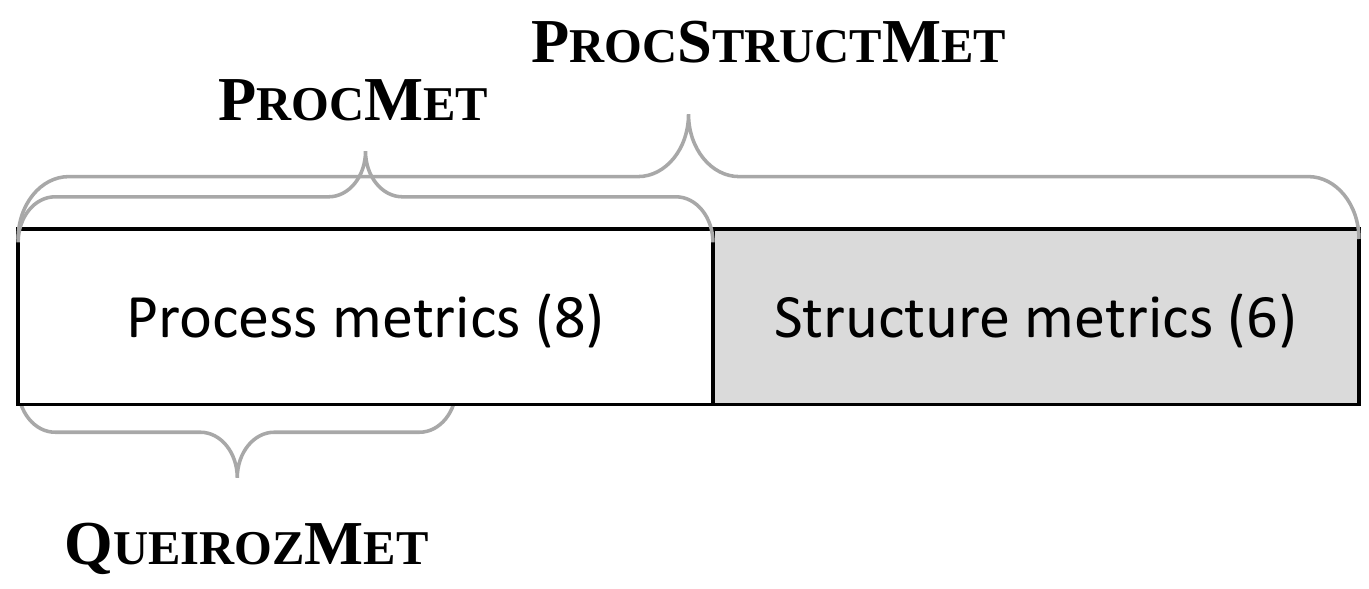}
	\vspace{-.4cm}
	\caption{Metric sets}
	\label{fig:metricsets}
\end{figure}

\begin{figure}[ht]
	\centering
	\includegraphics[width=\columnwidth]{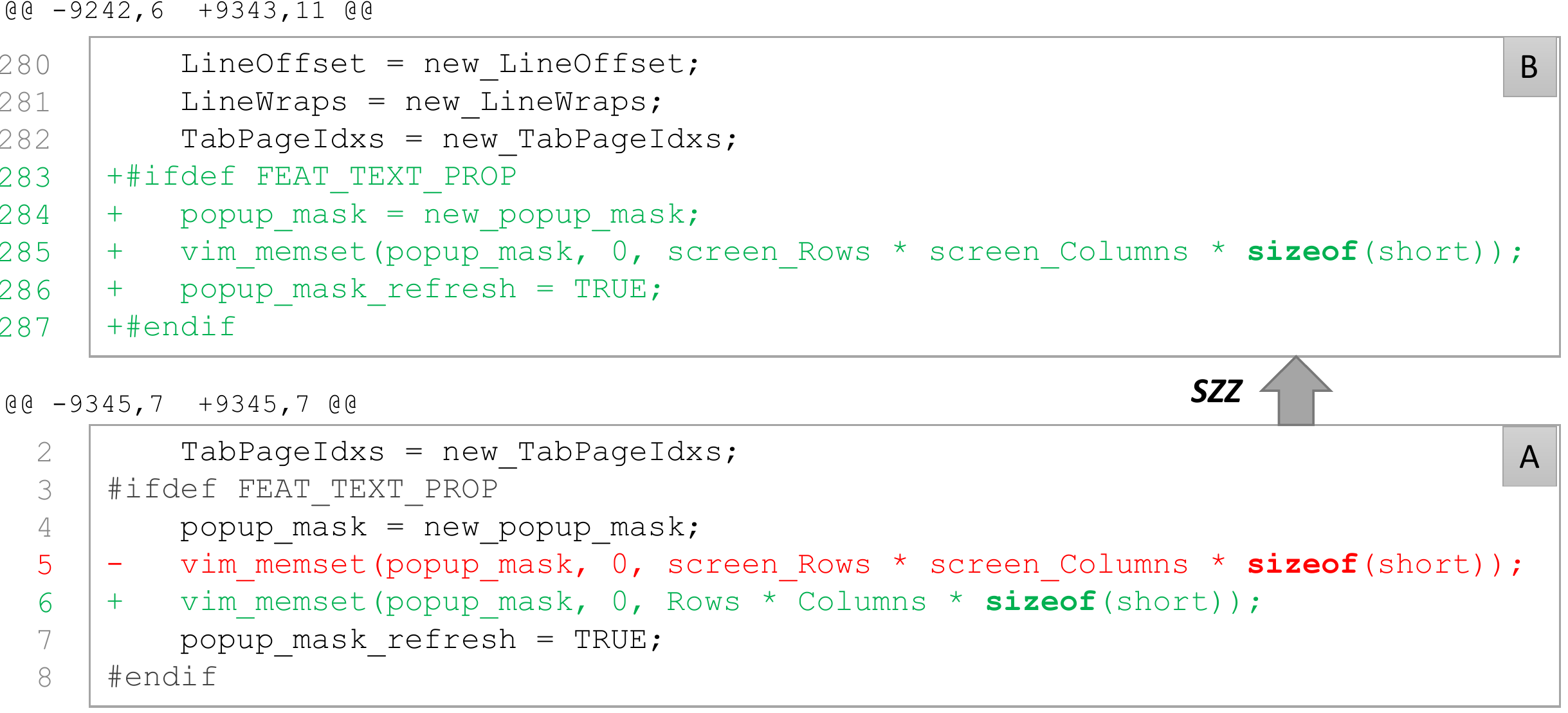}
	\vspace{-.7cm}
	\caption{Example of a defect with corrective (A) and bug-introducing (B) commit\label{fig:bug-example}}
	\vspace{-.4cm}
\end{figure}

\looseness=-1
\parhead{Metrics as attributes} 
To map the available feature information to attributes, we need to design a set of software metrics---numerical values that quantify properties of a software project.
We consider both \textit{structure metrics}, which are used to measure certain qualities of the software code of a specific revision, and \textit{process metrics},  which are used to measure properties of metadata taken from software repositories\,\cite{Rahman2013}, or take the evaluation of characteristics over revisions into account.
In the context of features, an example structure metric is: scattering degree (counting all preprocessor macro references to the feature---e.g., \hashifdef~\foperands{A}). An example process metric is: the number of committers who changed the feature in the release.

\parhead{Metric sets} 
We obtained three metric sets: an existing one from the literature (\metRod) and two new metric sets (\metA, \metB) we obtained by incrementally refining the available metrics. \metA\ augments \metRod\ with additional process metrics, whereas \metB\ extends \metA\ with structure metrics.
\autoref{fig:metricsets} illustrates the metric sets and their relationships.
\Tabref{tab:featureMetrics} gives a detailed overview of the resulting fourteen metrics and their descriptions.

\begin{itemize}[leftmargin=*]
	\item \textbf{\metRod}:
	The original metric set by \rodrigo\,\cite{Queiroz2016} consists of five process metrics, based on the rationale that process metrics are deemed particularly beneficial in defect prediction\,\cite{Rahman2013}.
	The included metrics quantify basic information such as the number of commits associated with the feature, developers contributing to the feature's implementation, and the experience of these developers (based on previous involvement).
	
	\item \textbf{\metA}: In  the first iteration, we systematically investigated additional process metrics.
	In the absence of dedicated feature process metrics in the literature	(see \secref{sec:relwork}), we derived three new ones from existing non-feature process metrics.
	These metrics quantify more involved feature-related process information, such as the average number of lines of code added to the files associated with the feature in the release.
	The original, file-based versions of these metrics were assessed as beneficial for defect prediction in earlier work\,\cite{Rahman2013}. 
	Consequently, we obtained a new metric set \metA, consisting of eight process metrics.
	
	\item \textbf{\metB}:
	In the second iteration, we systematically investigated feature structure metrics, aiming to benefit from two complementary types of metrics by bundling them.
	We added six structure metrics:
	four custom feature structure metrics identified in related work (see \secref{sec:relwork}), and two new ones we derived from particularly common structure metrics.
	The custom metrics include metrics such as the nesting depth of a feature (number of other preprocessor macros nested within the macro of a given feature). The newly derived metrics, based on LoC and cyclomatic complexity, represent the two main dimensions usually considered by structure metrics: size and complexity.
	Overall, our metric set\,\metB~ comprises eight process and six structure  metrics.
\end{itemize}

Generally, the metric values for each feature are aggregated over a release as described in \Tabref{tab:featureMetrics}. The values of the metrics are calculated for the data of each subject project, in some cases directly using SQL queries, in some cases by combining SQL queries and a Python script, and in other cases by using an available tool.

\setlength{\tymin}{3.5cm}
\begin{table*}[!t]
	\caption{Metrics  of metric sets \metA\ (process metrics) and \metB\ (process and structure metrics)}
	\fontsize{9}{9}\selectfont
	\vspace{-.1cm}
	\textbf{Legend}: Let $\mathnormal{R = \{c_{1}, c_{2}, . . . c_{q}}\}$ be a release set consisting of $q$ commits; $C$ be the set of all commits from previous releases plus those in $R$; $\mathnormal{F = \{f_{1}, f_{2}, . . . f_{p}}\}$ be the set of all files changed by commits in $R$. Let $\mathnormal{T = \{\textit{feat}_{1}, \textit{feat}_{2}, . . . \textit{feat}_{n}}\}$ be the set of all features affected by changes in $R$ (i.e., features included in diffs), where each feature $\textit{feat} {\in} T$ has a set $\mathnormal{A = \{\textit{featfile}_{1}, \textit{featfile}_{2}, . . . \textit{featfile}_{m}}\}$ of files implementing it, and $A\subseteq F$. We define our metrics for each feature \textit{feat}, with respect to release $R$ as follows:
	
	\tablabel{tab:featureMetrics}
	\begin{threeparttable}
		\fontfamily{ptm}\selectfont
		
		\begin{tabulary}{\textwidth}{@{}LLLL@{}}
			
			\toprule
			&\theader{metric\tnote{1}}&\theader{description}	&	\theader{function signature}	\\
			\midrule
			\multirow{3}{1.5mm}[-4mm]{\rotatebox{90}{process metrics}}
			&\metricName{FCOMM} & Count of all commits in which a feature was changed within a release. & $\mathnormal{comm(\feat,R)}$ \\			
			&\metricName{FADEV} & Count of all developers who changed a feature within a release. & $\mathnormal{adev(\feat,R)}$  \\
			
			&\metricName{FDDEV} & Count of all distinct developers who changed the feature up to the current release & $\mathnormal{ddev(\feat, C)}$   \\
			
			&\metricName{FEXP}\tnote{2} & Average experience\mtnote{3} of all developers who changed a feature within a release. & $ \mathnormal{ exp(\feat, R)}$ \\
			
			&\metricName{FOEXP} & Average experience of the developer who changed the features of a file most often within a release. & $ \mathnormal{oexp(\feat,R)}$   \\
			
			&\metricName{FMODD} & Average scattering degree of a feature in changesets within a release---counts number of \#ifdef references to a feature within each changeset and averages this over the release & $modd(\feat,R)$  \\
			
			&\metricName{FADDL} & Average number of lines of code added to the files associated with a feature within a release. & $addl(\feat,A)$ \\
			
			&\metricName{FREML} & Average number of lines of code deleted from the files associated with a feature within a release. & $reml(\feat,A)$   \\
			\midrule
			
			\multirow{3}{1.5mm}[-4mm]{\rotatebox{90}{structure metrics}}			
			&\metricName{FNLOC} & Average number of lines of code of the files associated with a feature within a release. & $ nloc(\feat,A)$  \\		
			&\metricName{FCYCO} & Average cyclomatic complexity of the files associated with a feature within a release. & $ cyco(\feat,A)$   \\
			&\metricName{LOFC} & Number of lines of code associated with a feature in a release (calculated from the last commit in $R$)  & \textit{lofc(\feat,$c_q$)}  \\
			
			&\metricName{NDEP} & Maximum nesting depth of \#ifdef directives that the feature is involved in (calculated from the last commit in $R$) & \textit{ndep((\feat,$c_q$)}   \\
			
			&\metricName{SCAT} & Scattering degree of a feature---count of all \#ifdef references to the feature (calculated from the last commit in $R$) & \textit{scat(\feat,$c_q$)}  \\
			&\metricName{TANGA} &Tangling degree of a feature---count of all other features mentioned the \#ifdef reference as the feature, e.g., \textit{\#ifdef featA \& featB} (calculated from the last commit in $R$) & \textit{tang(\feat,$c_q$)}  \\	
			\bottomrule				
		\end{tabulary}	
		\begin{tablenotes}[para]
			
			~\item[1] The first five process metrics (\metricName{FCOMM}, \metricName{FADEV}, \metricName{FDDEV}, \metricName{FEXP}, and \metricName{FOEXP}),with a slight modification in names, were introduced by \rodrigo\,\cite{Queiroz2016}; here we prefixed them with F to indicate that they are calculated over features unlike commonly done with files e.g., by Rahman et al. \,\cite{Rahman2013}. We refer to this set of metrics as \metRod.
			
			~\item[2] $exp(\feat, R)$ returns the geometric mean of the experience\mtnote{3} of all developers who changed the feature within a release.
			
			~\item[3] Experience is the sum of the changed, deleted or added lines in the commits associated with the files (set $A$) implementing \feat.
		\end{tablenotes}
	\end{threeparttable}
	\vspace{-0.3cm}
	
\end{table*}

\subsubsection{Selection and Training of Classifiers}
\label{sec:classifiers}

We selected seven classifiers based on their use in previous studies: J48 Decision Trees (J48), k-Nearest-Neighbors (KNN), Logistic Regression (LR), Na\"{\i}ve Bayes Bayes (NB), Artificial Neural Networks (NN), Random Forest (RF), and Support Vector Machines (SVM).     
A key informative work for our selection was the empirical study by Son et al.~\cite{Son2019}, who determine the six most commonly used classifier types in 156 defect-prediction studies:
Decision Tree, Random Forest, Bayesian, Regression, Support Vector Machines and Neural Networks.
We used typical representative learners for each of the broader categories: J48 (Decision Tree), LR (Regression), NB (Bayesian). As an  example for learners that are commonly used in classification, but less so in defect prediction, we included k-Nearest Neighbor (KNN).

\looseness=-1
\parhead{Tool and configuration} 
To train and test our classifiers, we used the WEKA workbench\footnote{\url{https://www.cs.waikato.ac.nz/ml/weka/}}, which is widely used in scientific studies, including defect prediction\,\cite{Hammouri2018,Ratzinger2008,Queiroz2016}. All our selected classification algorithms are already integrated in WEKA.

We trained each classification algorithm in WEKA with the respective standard settings, except for NN and RF where we set the number of decision trees for RF to 200, and used a hidden layer structure of \texttt{(13,13,13)} for NN for more efficient processing. We selected these settings independently since no specific recommendations exist.

We trained each of our seven classifiers using the dataset with each of the three  metric sets, leading to 21 instances of training in total. We used a Windows 10 system (Intel Core i7-6500U, 16GB RAM) for all experiments. The training times varied between a few seconds and slightly above a minute. NN recorded the longest time (52 sec with \metRod\, and 65 sec with \metB), while KNN recorded the shortest (0.01 sec with \metRod\, and 0.02 sec with \metB.)

\parhead{Test vs. training set} 
We determined the ratio of training data to test data for each individual project based on the number of available releases. We aimed to approximate the commonly used split ratios of between $80:20\%$ and $70:30\%$. The resulting ratio splits, as shown in \Tabref{tab:datasetexample}, range from $67:33\%$ to $80:20\%$.

In general, we assigned earlier releases to the training data and later ones to test data. In doing so, we avoid the implausible situation of "using the future to predict the past", which is unrealistic in practice\,\cite{jimenez2019importance}. For the same reason, we do not use cross-validation to evaluate our selected classifiers.

\parhead{Imbalanced dataset} 
\Tabref{tab:dataset-numbers} reveals that our dataset is imbalanced: \textit{clean} instances outnumber \textit{defective} ones by a factor of 4.14. Using imbalanced datasets for training is generally known to skew the classifier towards misclassification of the under-represented class. A common mitigation strategy is to apply over-sampling, by generating synthetic examples of the minority class. To this end, we apply the SMOTE\,\cite{Chawla2002} algorithm to our training dataset by using the available implementation in WEKA, in its standard configuration.

\subsection{Results}
\looseness=-1
Using our trained classifiers (\secref{sec:dataset}) and the three considered metric sets (\secref{sec:rq1method}), we studied three research questions:

\begin{itemize}[leftmargin=*]
	\item \textit{RQ1.1: What is the effect of using different types of feature metrics (structural and process) on prediction quality?}
	\item \textit{RQ1.2: Which particular feature metrics contribute most strongly to prediction quality?}
	\item \textit{RQ1.3: What is the effect of using different classifiers on prediction quality?}
	
\end{itemize}

Within RQ1, we implicitly compare our contribution to the most closely related work: our two new metric sets are compared to the one from \rodrigo\,\cite{Queiroz2016}, who proposed the only other dedicated metric set for feature-oriented defect prediction.

\Tabref{tab:evalmetrics} and \figref{roc-feat} together give an overview of our results.
\Tabref{tab:evalmetrics} provides all precision, recall, F-score, and AUC values.
For each classifier and evaluation metric, the top value (best-perfoming metric set) is highlighted in bold.
\Figref{roc-feat} shows ROCs for three representative classifiers (top performer, average performer, worst performer in terms of AUC) in combination with all three metrics sets.
For reference, AUC values of all cases are shown in the table.

\begin{table*}
	
	\centering
	\caption{Results RQ1 and RQ3: evaluation metrics for the classes ``defective'' and ``clean,'' and the weighted average ``w.a.''}
	\vspace{-.2cm}
	\tablabel{tab:evalmetrics}
	\fontsize{8}{8}\selectfont
	\begin{tabular}{llrrrrrrrrrrr} 
		\toprule
		&                  & \multicolumn{11}{c}{\textbf{Metric set} } \\[+.1cm] 
		&                  & \multicolumn{3}{c}{\metRod} & & \multicolumn{3}{c}{\metA} & & \multicolumn{3}{c}{\metB}  \\
		\cline{3-5}
		\cline{7-9}
		\cline{11-13}\\
		\textbf{Classifier}                    & \textbf{Eval. metric}                 & defective & clean  & w.a.  &    & defective & clean  & w.a.  &  & defective & clean  & w.a.                \\ 
		\midrule
		\multirow{4}{*}{J48} & Recall & 0.57    & 0.66 & 0.64 &   & 0.61    & \textbf{0.85} & 0.80 &&  \textbf{0.65}    & \textbf{0.85} & \textbf{0.81}              \\
		& Precision        & 0.27    & 0.87 & 0.77   && 0.47    & 0.91 & 0.83 && \textbf{0.49}    & \textbf{0.92} & \textbf{0.84}              \\
		& F-score          & 0.37    & 0.75 & 0.68   && 0.53    & \textbf{0.88} & 0.81 & & \textbf{0.56}   & \textbf{0.88} & \textbf{0.82}              \\
		& AUC area         & 0.57    & 0.57 & 0.57   && \textbf{0.79}    & \textbf{0.79} & \textbf{0.79} && 0.78    & 0.78 & 0.78              \\ 
		\midrule
		\multirow{4}{*}{KNN} & Recall & 0.53    & 0.56 & 0.55   && \textbf{0.57}    & 0.58 & 0.58 && 0.55    & \textbf{0.81} & \textbf{0.77}              \\
		& Precision        & 0.21    & 0.84 & 0.73   && 0.23    & 0.86 & 0.75 && \textbf{0.39}    & \textbf{0.89} & \textbf{0.80}              \\
		& F-score          & 0.30    & 0.67 & 0.61   && 0.33    & 0.69 & 0.63 && \textbf{0.46}    & \textbf{0.85} & \textbf{0.78}              \\
		& AUC area         & 0.50    & 0.50 & 0.50   && 0.52    & 0.52 & 0.52 && \textbf{0.74}    & \textbf{0.74} & \textbf{0.74}              \\ 
		\midrule
		\multirow{4}{*}{LR}  & Recall & 0.40    & \textbf{0.73} & \textbf{0.67}   && 0.43    & 0.72 & \textbf{0.67} && \textbf{0.45}    & 0.72 & \textbf{0.67}              \\
		& Precision        & 0.25    & \textbf{0.85} & 0.74   && 0.25    & \textbf{0.85} & 0.74 && \textbf{0.26}    & \textbf{0.85} & \textbf{0.75}              \\
		& F-score          & 0.30    & \textbf{0.78} & \textbf{0.70}   && 0.32    & \textbf{0.78} & \textbf{0.70} && \textbf{0.33}    & \textbf{0.78} & \textbf{0.70}              \\
		& AUC area         & \textbf{0.64}    & \textbf{0.64} & \textbf{0.64}   && 0.60    & 0.60 & 0.60 && 0.60    & 0.60 & 0.60              \\ 
		\midrule
		\multirow{4}{*}{NB}  & Recall & 0.38    & \textbf{0.94} & \textbf{0.84}   && \textbf{0.40}    & 0.93 & \textbf{0.84} & & 0.37    & \textbf{0.94} & \textbf{0.84}              \\
		& Precision        & \textbf{0.58}    & 0.87 & \textbf{0.82}   && 0.57    & \textbf{0.88} & \textbf{0.82} && 0.57    & 0.87 & \textbf{0.82}              \\
		& F-score          & \textbf{0.50}    & \textbf{0.91} & \textbf{0.82}   && 0.47    & 0.90 & \textbf{0.82} && 0.45    & 0.90 & \textbf{0.82}              \\
		& AUC area         & 0.61    & 0.61 & 0.61   && 0.77    & 0.77 & 0.77 && \textbf{0.78}    & \textbf{0.78} & \textbf{0.78}              \\ 
		\midrule
		\multirow{4}{*}{NN}  & Recall & 0.28    & 0.75 & 0.66&   & 0.30    & 0.75 & 0.67 && \textbf{0.33}    & \textbf{0.97} & \textbf{0.85}              \\
		& Precision        & 0.20    & 0.82 & 0.71   && 0.21    & 0.83 & 0.72 && \textbf{0.69}    & \textbf{0.87} & \textbf{0.84}              \\
		& F-score          & 0.23    & 0.78 & 0.68   && 0.25    & 0.79 & 0.69 && \textbf{0.45}    & \textbf{0.92} & \textbf{0.83}              \\
		& AUC area         & 0.55    & 0.55 & 0.55   && 0.61    & 0.61 & 0.61 && \textbf{0.79}    & \textbf{0.79} & \textbf{0.79}              \\ 
		\midrule
		\multirow{4}{*}{RF}  & Recall & 0.57    & 0.63 & 0.62   && 0.62    & 0.83 & 0.80 && \textbf{0.68}    & \textbf{0.85} & \textbf{0.82}              \\
		& Precision        & 0.26    & 0.87 & 0.76   && 0.45    & 0.91 & 0.83 && \textbf{0.51}    & \textbf{0.92} & \textbf{0.85}              \\
		& F-score          & 0.35    & 0.73 & 0.66   && 0.52    & 0.87 & 0.81 && \textbf{0.58}    & \textbf{0.89} & \textbf{0.83}              \\
		& AUC area         & 0.59    & 0.59 & 0.59   && 0.75    & 0.75 & 0.75 && \textbf{0.82}    & \textbf{0.82} & \textbf{0.82}              \\ 
		\midrule
		\multirow{4}{*}{SVM} & Recall & 0.12    & \textbf{1.00} & \textbf{0.84}   && 0.22    & 0.76 & 0.66 && \textbf{0.23}    & 0.76 & 0.66              \\
		& Precision       & \textbf{0.83}    & \textbf{0.84} & \textbf{0.84}   && 0.17    & 0.82 & 0.70 && 0.17    & 0.82 & 0.70              \\
		& F-score          & \textbf{0.21}    & \textbf{0.91} & \textbf{0.78}   && 0.19    & 0.79 & 0.68 && 0.20    & 0.79 & 0.68              \\
		& AUC area         & \textbf{0.56}    & \textbf{0.56} & \textbf{0.56}   && 0.49    & 0.49 & 0.49 && 0.49    & 0.49 & 0.49              \\
		\bottomrule
	\end{tabular}
	\vspace{-.3cm}
\end{table*}

\begin{figure*}[t]
  \centering
  \subfloat[][RF]{\includegraphics[width=0.33\linewidth]{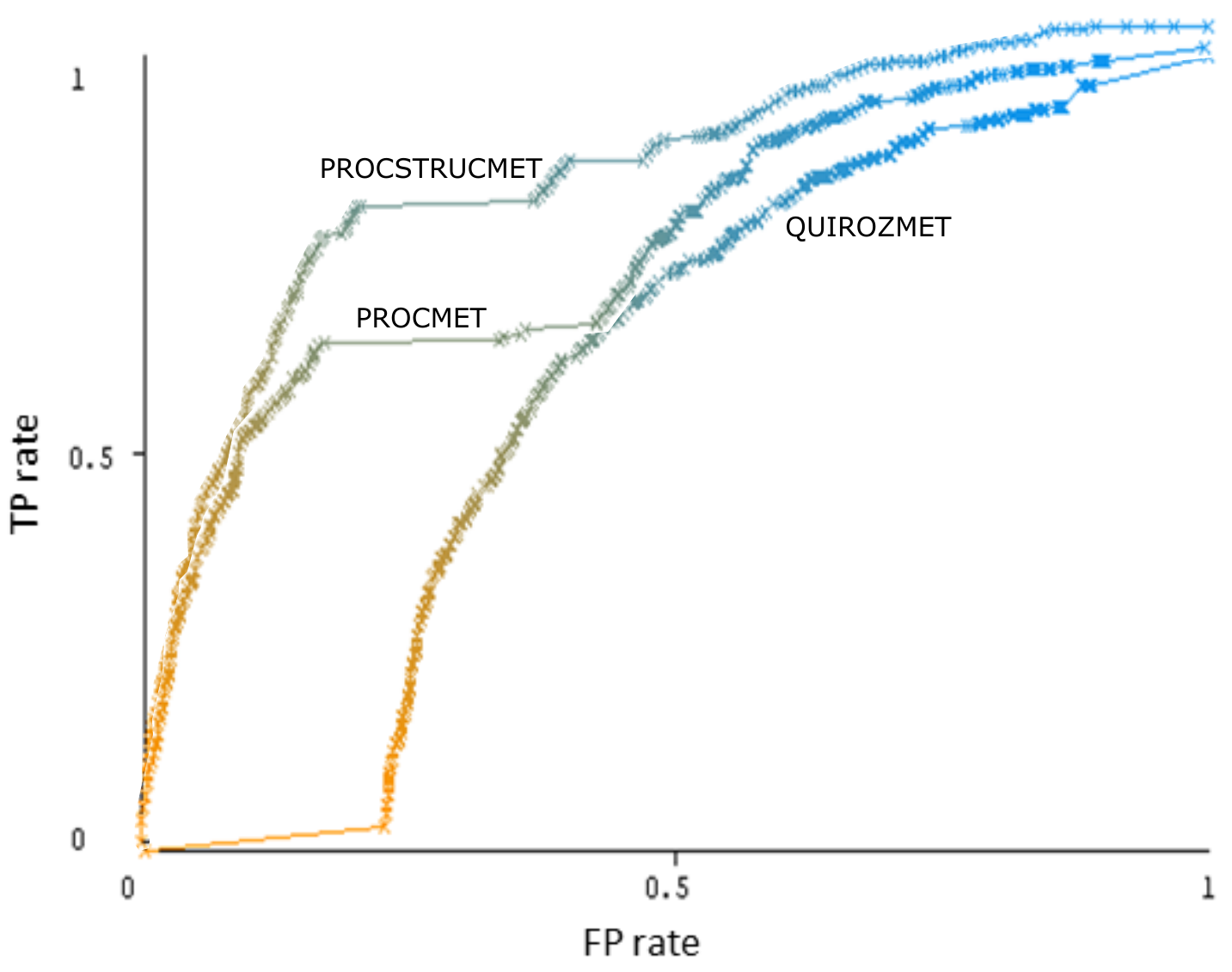}} 
  \subfloat[][KNN]{\includegraphics[width=0.33\linewidth]{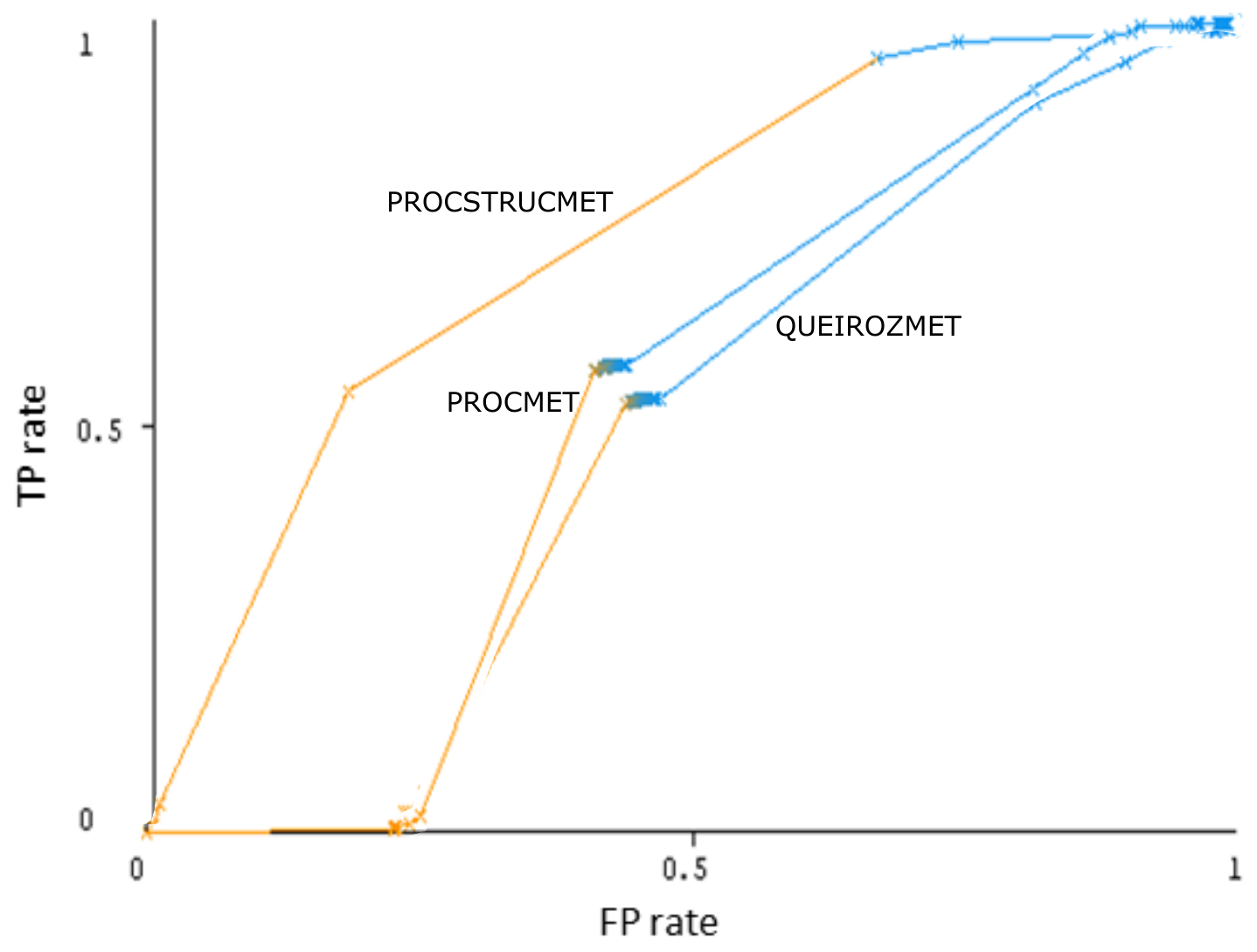}}
  \subfloat[][SVM]{\includegraphics[width=0.33\linewidth]{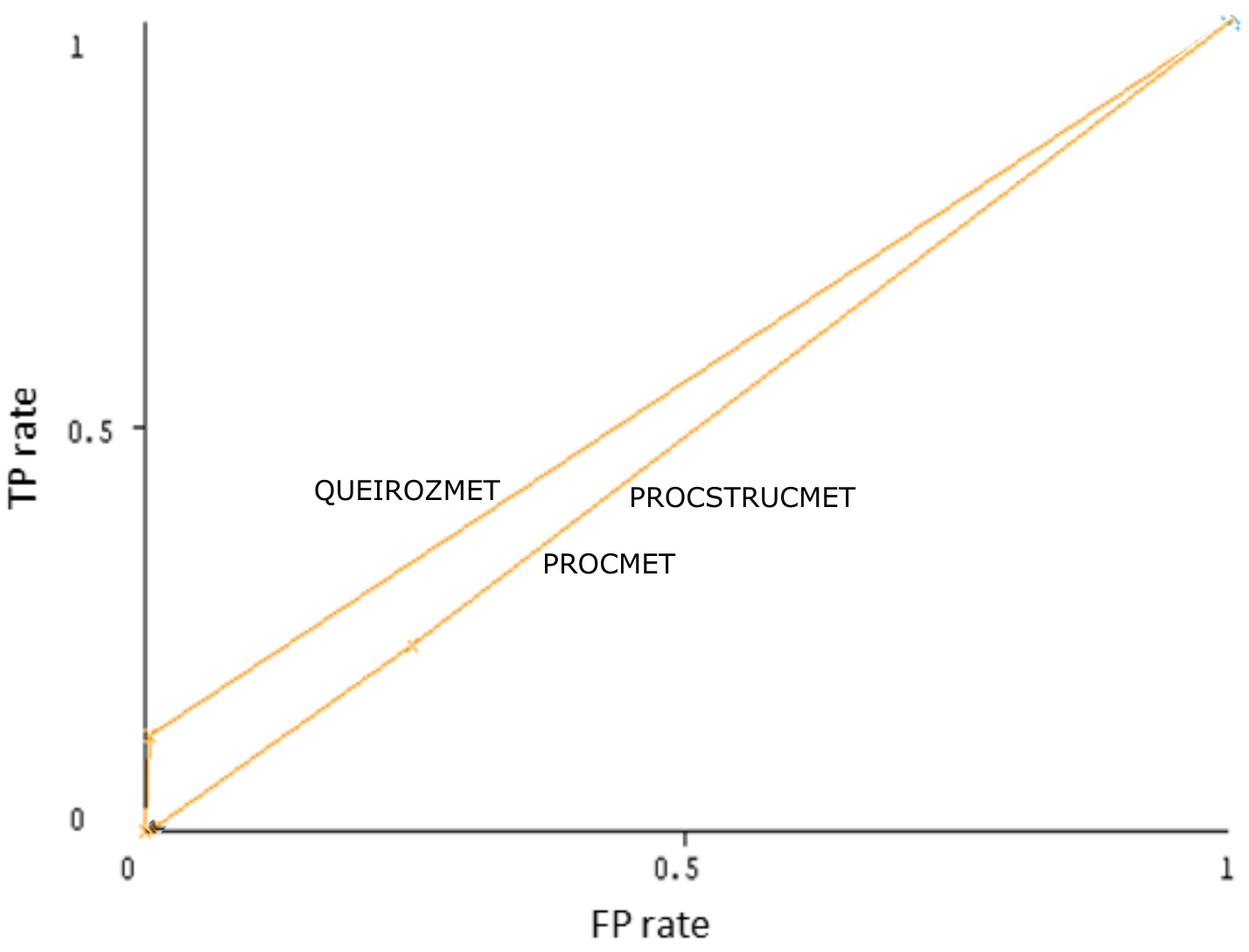}}
	\vspace{-.2cm}
  \caption{Results RQ1.1 and RQ1.3: ROC curves for three selected classifiers (top, average, worst) \label{fig:roc-feat}}
\vspace{-.3cm}
\end{figure*}

\begin{table}[b]
\vspace{-.3cm}
\centering
\caption{Results RQ1.2: Influence of metrics for top classifiers (RF, NN, J48 with \metB) and all classifiers}
\label{tab:eval-metrics}
\vspace{-.2cm}
\begin{tabular}{lllllll}
			\toprule				
&      &       &        &       & \multicolumn{2}{c}{\textbf{All classifiers}}\\ 
\cline{6-7} & \textbf{Metric}     & \textbf{RF}    & \textbf{NN}     & \textbf{J48}   & mean          & std.dv          \\ 
			\midrule				
\multirow{8}{0.5mm}[-1mm]{\rotatebox{90}{process metrics}}%
			& \metricName{FCOMM} & 0.048 & 0      & \textbf{0.049} & \textbf{0.034}              & 0.02            \\ 
&\metricName{FADEV} & 0.025 & \textbf{0.024}  & 0.025 & 0.022              & 0.004           \\ 
&\metricName{FDDEV} & 0.009 & 0.008  & 0.01  & 0.009              & 0.003           \\ 
&\metricName{FEXP}  & 0.047 & 0      & 0.018 & 0.006              & 0.036           \\ 
&\metricName{FOEXP} & 0.057 & 0      & 0.036 & 0.021              & 0.026           \\ 
&\metricName{FMODD} & 0.042 & 0      & 0.041 & 0.028              & 0.019           \\ 
&\metricName{FADDL} & \textbf{0.058} & 0      & 0.035 & -0.03              & 0.142           \\ 
&\metricName{FREML} & 0.03  & 0      & 0.016 & 0.011              & 0.014           \\ 
			\midrule	
\multirow{6}{1.5mm}[0mm]{\rotatebox{90}{structure metrics}}%
			&\metricName{FNLOC} & 0.014 & 0      & 0.002 & 0.004              & 0.006           \\ 
&\metricName{FCYCO} & 0.013 & 0      & 0.007 & 0.004              & 0.005           \\ 
&\metricName{LOFC}  & 0.002 & 0      & 0     & 0                  & 0.001           \\ 
&\metricName{NDEP}  & 0.002 & 0      & 0     & 0                  & 0.001           \\ 
&\metricName{SCAT}  & 0.005 & -0.001 & 0     & 0.001              & 0.002           \\ 
&\metricName{TANGA} & 0     & 0      & 0     & 0                  & 0               \\ 
			\bottomrule				
\end{tabular}
\end{table}

\parhead{RQ1.1: Effect of metric sets} 
Based on precision, recall, and F-score, we generally observe a moderate tendency that classifiers performed best when using \metB, for which we observed weighted averages between 0.66--0.85, 0.70--0.85 and 0.68--0.83 respectively.
The corresponding ranges for the case of  \metA\ and \metRod\ are 0.58--0.84, 0.70--0.83 and 0.63--0.82,
and  0.55--0.84, 0.71--0.84 and 0.61--0.82, respectively.
The quality difference is particularly pronounced when considering the top values (printed in bold):
In all classifiers except for SVM, \metB\ shows the top value for precision, recall, and F-score.
Two noteworthy observations are the case of NB, where all evaluation metrics take the same values over all metrics sets, and SVM, where \metRod\ outperforms \metA\ and \metB.
Considering the two classes \textit{clean} and \textit{defective}, we generally find higher F-scores in the more advanced metric sets, and a better ability to predict clean than defective instances for all metric sets.

\looseness=-1
Considering ROCs and AUCs sheds light on the effect of the metric sets on robustness.
We generally find a clear tendency of \metB\ to highest robustness, i.e., more stability with regard to different values for the threshold used for assigning instances to classes (reflected by a steeper initial incline in the ROC curves).
In 5 out of 7 cases, the AUC for \metB\ shows a solid value between 0.74 and 0.82.
The AUC for \metB\ is consistently greater or equal to that of \metA\, in some cases strongly so, including the top performer NN (0.79 vs. 0.61).
The highest achieved value for \metRod\ is 0.64.
The SVM classifier is an exception to all other cases: for \metA\ and \metB\, we observe worse performance (0.49) than from random guessing (0.5); corresponding to a nearly-linear ROC. A possible explanation for the preferable robustness of \metB\ in most classifiers is the availability of more diverse metrics, providing a richer information source for predictions.

\looseness=-1
\parhead{RQ1.2: Effect of individual metrics} 
We determined the effect of individual metrics with an attribute selection method.
Such methods heuristically determine the effect of attributes (i.e., our metrics) with regard to a classifier's predictive ability.
We used a standard method provided by Weka (\texttt{weka.attributeSelection. ClassifierAttributeEval} together with the \texttt{Ranker} class).
This method runs the considered classifier several times with different subsets of the entire metric set and outputs an influence measure between 1.0 and -1.0 for each metric, quantifying its influence on the prediction result.
We applied the method to all 21 classifier instances (7 classifiers with 3 metric sets).

We present an overview of the results in \autoref{tab:eval-metrics}, showing the three top performers from RQ1 and the 
average over all 21 classifier instances.
The most influential metric for each classifier is highlighted in bold.
Generally, the obtained values are very similar for RF and J48, perhaps unsurprisingly, since RF and J48 are both based on the decision tree paradigm.
For NN, only three non-zero values are reported, which, however, agree with the reported values for the other two top classifiers.
We observe striking cases of large standard deviations, most pronounced in the case of FADDL, which has the most positive impact for the RF classifier (0.058), while, on average, leading to a strong negative influence (-0.03).
Despite the observation in RQ1 that the inclusion of structure metrics leads to improved results compared to only process metrics, the effect of each individual structure metric is moderate compared to the process metrics.
This indicates that structure metrics seem to play a non-negligible, but supplementary role for the observed results.

\parhead{RQ1.3: Effect of classifiers} 
\looseness=-1
As a general observation, in most cases, the prediction quality of the same classifier varied strongly based on the considered metrics set (see RQ1).
It is, therefore, more meaningful to compare combinations of classifiers and metrics, rather than classifiers alone.
Considering weighted averages, we observe values for precision between 0.70--0.85, for recall between 0.62--0.85, and for F-score between 0.61--0.83.
The best-performing classifiers with regard to F-score were NN and RF, both in combination with \metB\ showing an F-score of 0.83.
NN also shows the best average-weighted recall (0.85), while RF shows the top average-weighted precision (0.85).
With a value of 0.82 for \metB, J48 also achieved above-average performance.

\looseness=-1
Considering individual classes, similar to the comparison between metrics, the results for the label \textit{defective} are generally worse than those of the label \textit{clean}.
The best precision for predicting clean files, 0.83, is observed for SVM in combination with \metRod.
However, this value is traded off for the worst observed recall for that label (0.12).
Note that the seemingly contra-intuitive average score of 0.78 for this combination results from averaging over the individual F-scores for class labels (0.21 and 0.91).

\looseness=-1
Considering ROC curves and AUC areas, we again observe a large inter-classifier variability.
Still, the two top classifiers with regard to precision, recall, and F-balance also have the two highest observed AUC values:
RF with 0.82, and NN with 0.79, indicating that these classifiers a good robustness while ensuring high predictive ability. An interesting observation is that the minimal AUC value achieved per classifier was never higher than 0.61, close to random guessing. In contrast, in four out of seven cases, a maximal value of 0.78 could be observed -- illustrating again that the choice of metric set is a key decisive factor for the success or failure of a particular predictor.

\looseness=-1
\parhead{Summary} 
We find a strong effect of metric selection on the classifier performance.

\insight{In most cases, we notice that considering a greater selection of more diverse metrics (that we introduce in this paper) lead to improved performance, including the identified top performers NN and RF.
However, this tendency does not apply to all cases: a remarkable counterexample are SVMs, where the predictive ability declines with the availability of more metrics. Hence, identifying a metric set that improves the performance of the considered classifier appears to be a key prerequisite to successful adoption of machine learning techniques for feature-based defect prediction.}

\setlength{\tymin}{3.5cm}
\begin{table*}[!t]
	\centering
	\caption{Ranking of file+feature-based metrics combined for file defect prediction}
	\fontsize{9}{9}\selectfont
	\tablabel{tab:rq2MetricRankingResults}
	\begin{threeparttable}
		\fontfamily{ptm}\selectfont
		
		\begin{tabulary}{\textwidth}{@{}LCCCCCCCCCCCCCCCC@{}}
			
			\toprule
			rank&1&2&3&4&5&\textbf{6}&\textbf{7}&\textbf{8}&\textbf{9}&\textbf{10}&\textbf{11}&\textbf{12}&\textbf{13}&\textbf{14}&\textbf{15}&\textbf{16}\\
			\midrule
			metric&avgc&maxc&aage&wage&auth&\textbf{fadev}&\textbf{fddev}&\textbf{scat}&\textbf{fnloc}&\textbf{fcyco}&\textbf{fmodd}&\textbf{fcomm}&\textbf{foexp}&\textbf{tanga}&\textbf{ndep}&\textbf{lofc}\\
			\midrule
			rank&17&\textbf{18}&\textbf{19}&20&21&\textbf{22}&23&24&25&\textbf{26}&27&28&29&30&31&32\\
			\midrule
			metric&bugf&\textbf{freml}&\textbf{fexp}&refa&cchm&\textbf{faddl}&adda&addm&revi&\textbf{fnof}&ccha&rema&remm&reml&addl&cchl		\\

			\bottomrule				
		\end{tabulary}	
		\begin{tablenotes}[para]

		\end{tablenotes}
	\end{threeparttable}
	\vspace{-0.3cm}
\end{table*}

\section{Effect of Feature-based Metrics on File-based Defect Prediction (RQ2)}
\seclabel{sec:rq2effectfilebasedmetrics}
\looseness=-1
We now address RQ2 (\textit{What is the effect of feature-based metrics on file-level defect prediction?}), describing methodology and results.
 
\subsection{Methodology} 
Similar to our feature-based dataset used to address RQ1, we created a release-level file-based dataset using 17 metrics (see \tabref{tab:fileMetrics}) from Moser et al.\,\cite{Moser2008}.  We then split the dataset into training and test datasets using the proportions in \tabref{tab:datasetexample} (the same way we did for the feature-based dataset, as explained in \secref{sec:rq1method}). We ran the seven classifiers on the file-based dataset to determine the best performing classifier and found that Random Forest outperformed the others, with J48 being the second best. Therefore, we only used Random Forest to answer RQ2-RQ5.

To apply feature-based metrics (\tabref{tab:featureMetrics}) to each file in the file-based dataset, we proceeded as follows: Given a file \textit{file1} that has been changed in a release \textit{R} and contains code implementing features \textit{feat1}, \textit{feat2}, and \textit{feat3}: we calculated metrics for each of the features in \textit{file1} as described in \tabref{tab:fileMetrics}. For each metric, we get the maximum value from all values returned for each feature.  For instance, for the metric \texttt{\textsc{FCOMM}} that counts commits in which a feature was modified, if \textit{fcomm(feat1,R)}=2,  \textit{fcomm(feat2,R)}=3, and \textit{fcomm(feat3,R)}=5, then \textit{fcomm(file1,R)}=5, indicating that the features in \textit{file1} changed a maximum of 5 commits in the release R. An alternative could have been to use the average value of the metrics instead of maximum. In addition to the 14 feature-based metrics, we added one more (\metricName{fnof}) to count the number of features in a file.  Thus, the dataset with combined metrics had in total 32 metrics; 17 file-based (\tabref{tab:fileMetrics}) and 15 feature-based (14 in \tabref{tab:featureMetrics} plus \metricName{fnof}).

We then evaluated the performance of the Random Forest classifier on the two datasets: the first with file-based metrics only, and the second with feature-based metrics included. We also used \textit{ReliefF} to rank the metrics for importance in the two datasets. The ranking allowed us to add two variations of each dataset, one with the top 75\% metrics and the other with the top 50\,\% metrics, to assess the best performing set of metrics.

\subsection{Results}
\seclabel{sec:rq2results}
\Tabref{tab:rq2SummaryResults} summarizes the results. \insight{We found the best performance (ROC 74.6\,\%) when using the top 75\,\% of metrics in the combined file+feature-based dataset. However, the difference in performance is marginal (file-only metrics gave ROC value of 72.2\,\%)}

Despite the marginal difference in performance, all feature-based metrics but \metricName{fnof} were in the top 75\,\% of the metrics for the combined dataset, and half of the top 10 metrics were feature-based. \Tabref{tab:rq2MetricRankingResults} shows the ranking of all 32 combined metrics; the feature-based metrics are highlighted in bold. 

The top ten ranking reveals that for any given file in our dataset, the number of files committed together with the file (\metricName{avgc,maxc}), the age of the file (\metricName{aage, wage}), and the number of authors of the file (\metricName{auth}), constitute the top 5 file-based metrics. Furthermore, the number of developers of features in the file (\metricName{fadev,fddev}), the scattering degree of features in the file (\metricName{scat}), the number of lines of code associated with features in the file (\metricName{fnloc}), and the cyclomatic complexity of features in the file (\metricName{fcyco}), are the top 5 feature-based metrics for predicting the defect-proneness of the file, ranking 6th to 10th respectively. Interestingly, feature-based metrics dominate the top 50\,\% ranking positions (6-16) for best predictors of defect proneness of files.

Therefore, we observe that there is a bigger spread of effects for the file-based metrics than the feature-based ones. The highest and least impact on prediction performance was by file-based metrics.

\setlength{\tymin}{3.5cm}
\begin{table}[tb]
	\centering
	\caption{Results summary: applying feature metrics to files}
	\fontsize{8}{8}\selectfont
	
	\tablabel{tab:rq2SummaryResults}
	\begin{threeparttable}
		\fontfamily{ptm}\selectfont
		
		\begin{tabulary}{\textwidth}{@{}LLLL@{}}
			
			\toprule
			\theader{file-based only}&\theader{ROC}	&	\theader{file+feature-based}&\theader{ROC}	\\
			\midrule			
				 all 17 metrics & 0.722 &all 32 metrics & 0.737 \\
			top 75\,\% (13 metrics)&0.737&top 75\,\% (24 metrics)&\textbf{0.746}\\
			top 50\,\% (9 metrics)&0.715&top 50\,\% (16 metrics)& 0.703\\
			\bottomrule				
		\end{tabulary}	
		\begin{tablenotes}[para]

		\end{tablenotes}
	\end{threeparttable}
	\vspace{-0.3cm}
	
\end{table}

\setlength{\tymin}{3.5cm}
\begin{table}[b]
\caption{Process metrics for file-based defect prediction \,\cite{Moser2008}}
	\fontsize{7}{7}\selectfont
	
	\tablabel{tab:fileMetrics}
	\begin{threeparttable}
		\fontfamily{ptm}\selectfont
		
		\begin{tabulary}{\columnwidth}{@{}LL@{}}
			\toprule
			\theader{metric}&\theader{description}		\\
			\midrule
			
			\metricName{REVISIONS (revi)} & Number of revisions of a file.  \\			
			\metricName{REFACTORINGS (refa)} & Number of times a file has been refactored.   \\
			
			\metricName{BUGFIXES (bugf)} & Number of times a file was involved in bug-fixing.    \\
			
			\metricName{AUTHORS (auth)} & Number of distinct authors that checked a file into the repository.  \\
			
			\metricName{LOC\_ADDED (addl)} & Sum over all revisions of the lines of code added to a file.    \\
			
			\metricName{MAX\_LOC\_ADDED (addm)} & Maximum number of lines of code added for all revisions.   \\
			
			\metricName{AVE\_LOC\_ADDED (adda)} & Average lines of code added per revision.  \\
			
			\metricName{LOC\_DELETED (reml)} & Sum over all revisions of the lines of code deleted from a file.   \\
			
			\metricName{MAX\_LOC\_DELETED (remm)} & Maximum number of lines of code deleted for all revisions.   \\		
			\metricName{AVE\_LOC\_DELETED (rema)} & Average lines of code deleted per revision.    \\
			\metricName{CODECHURN (cchn)} & Sum of added and deleted lines of code over all revisions.     \\
			
			\metricName{MAX\_CODECHURN (cchm)} & Maximum CODECHURN for all revisions.    \\
			
			\metricName{AVE\_CODECHURN (ccha)} & Average CODECHURN per revision.   \\
			\metricName{MAX\_CHANGESET (maxc)} &Maximum number of files committed together to the repository.   \\	
			\metricName{AVE\_CHANGESET (avgc)} & Average number of files committed together to the repository.   \\
			\metricName{AGE (aage)} &Age of a file in weeks (counting backwards from a specific release).   \\
			\metricName{WEIGHTED\_AGE (wage)} &$Weighted Age = \frac{\sum_{i=1}^N Age(i)*LOC\_ADDED(i)}{\sum_{i=1}^N LOC\_ADDED(i)}$   \\	
			\bottomrule				
		\end{tabulary}	
		\begin{tablenotes}[para]

		\end{tablenotes}
	\end{threeparttable}
	\vspace{-0.3cm}
	
\end{table}

\setlength{\tymin}{3.5cm}
\begin{table}[tb]
	\centering
	\caption{Summary of result for feature-to-file mapping with predicted labels}
	\fontsize{7}{7}\selectfont
	\tablabel{tab:rq3Results}
	\begin{threeparttable}
		\fontfamily{ptm}\selectfont
		
		\begin{tabulary}{\columnwidth}{@{}LCCCCCCC@{}}
			
			\toprule
			\theader{project}&	\theader{release}&	\theader{defective features}&	\theader{predicted}&	\theader{perc.\%}\tnote{1}&	\theader{defective files}&	\theader{predicted}&	\theader{perc.\%}\\
			\midrule
		blender&2.78&49&40&82\%&88&0&0\%\\
		&2.79&18&12&67\%&34&0&0\%\\
		&2.80&30&11&37\%&37&0&0\%\\
		total&&97&63&65\%&159&0&0\%\\
		\midrule
		busybox&1\_26\_0&11&2&18\%&16&0&0\%\\
		&1\_27\_0&3&1&33\%&3&0&0\%\\
		&1\_29\_0&2&2&100\%&2&0&0\%\\
		&1\_30\_0&1&1&100\%&1&0&0\%\\
		total&&17&6&35\%&22&0&0\%\\
			\midrule
		emacs&26.1&3&2&67\%&3&0&0\%\\
			\midrule
		gimp&2\_10\_10&1&0&0\%&1&0&0\%\\
		&2\_10\_8&1&1&100\%&1&0&0\%\\
		total&&2&1&50\%&2&0&0\%\\
			\midrule
		gnumeric&1\_12\_20&1&1&100\%&1&0&0\%\\
			\midrule
		gnuplot&5.0.0&40&20&50\%&80&0&0\%\\
			\midrule
		lighttpd&1.4.30&30&13&43\%&44&0&0\%\\
		&1.4.40&49&32&65\%&63&0&0\%\\
		&total&79&45&57\%&107&0&0\%\\
			\midrule
		parrot&6\_0\_0&51&35&69\%&57&0&0\%\\
			\midrule
		vim&8.0&52&30&58\%&75&0&0\%\\
		&8.1&77&71&92\%&122&0&0\%\\
		total&&129&101&78\%&197&0&0\%\\
			\midrule
		Grand Total&&419&274&65\%&628&0&0\%\\
			\bottomrule				
		\end{tabulary}	
		\begin{tablenotes}[para]
			
			\item[1] in the total row, percentages are averaged, not summed
		\end{tablenotes}
	\end{threeparttable}
	\vspace{-0.1cm}
\end{table}

\section{Feature-Based vs File-Based Defect Prediction (RQ3)}
\seclabel{sec:featureVsFilePredictions}
We now discuss RQ3 (\textit{How does feature-based defect prediction perform compared to file-based defect prediction?}).

\subsection{Methodology}
\seclabel{sec:rq3method}
Comparing the results from RQ1 and RQ2, we observed better better performance values for feature-based than for file-based defect prediction: 82\% for the former and 74.6\% for the latter. However, these results are not directly comparable since they concern different granularity levels (feature- and file granularity, respectively).
To allow a comparison, we used the file granularity as a common baseline, and studied whether feature-based or file-based defect prediction yields a higher number of correctly predicted defective files than the other. We proceeded as follows:
\begin{itemize}
	\item [1] We first mapped features to files that implement them. These mappings were determined during the data extraction phase (\secref{sec:dataset}).
	\item [2] Next, for both the file-based and feature-based test datasets, we generated corresponding CSV files containing all test-data (files or features) in the same order they appear in the ARFF test dataset files. For the data in the CSV files, we included the project name and feature/file name, in addition to the metric values and the label (defective or clean). Hence, the ARFF file and the CSV file had the same number of datapoints and in the same order.
	\item [3] Next, for both the feature-based and file-based datasets, we trained Random Forest (using the best performing set of metrics in the case of file-based dataset i.e. top 75\,\%, see \secref{sec:rq2results}) then predicted for each datapoint (file or feature) in the test dataset. For each test datapoint, we recorded its predicted label in the CSV file generated in step 2. Since the ARFF file and CSV file had datapoints in the same order, we relied on this to identify predicted labels for files or features in the respective CSV file. At the end of this step, the two CSV files (one for files test data and the other for features test data) had an extra column containing the predicted label for each file or feature.
	\item[4] Using the mapping of features to files generated in step 1, we generated one combined CSV file that mapped the features in the feature-based test dataset to their implementation files in the file-based test-dataset CSV file. \Tabref{tab:rq3featFileMapping} illustrates the format of the mapped CSV file. 
	\item[5] We then counted all cases where, for each defective feature correctly predicted as defective by the feature-based classifier, none of its mapped defective files are correctly predicted as defective by the file-level classifier. We also considered the opposite case where defective files are correctly predicted but none of the corresponding features are.
\end{itemize}

\setlength{\tymin}{3.5cm}
\begin{table}[tb]
	\centering
	\caption{Example result of feature-to-file mapping with predicted labels}
	\fontsize{8}{8}\selectfont
	
	\tablabel{tab:rq3featFileMapping}
	\begin{threeparttable}
		\fontfamily{ptm}\selectfont
		
		\begin{tabulary}{\columnwidth}{@{}LLLLLLLL@{}}
			
			\toprule
			\theader{project}&	\theader{release}&	\theader{feature}\tnote{1}&	\theader{feat. label}&	\theader{pred. feat. label}&	\theader{file}\tnote{1}&	\theader{file label}&	\theader{pred. file label}\\
			\midrule
			blender&2.78&\_ao\_&def.&def.&kernpb.h&def.&clean\\
			blender&2.78&\_ao\_&def.&def.&kernp.h&def.&clean\\
			blender&2.78&backg&def.&def.&kernpb.h&def.&clean\\
			blender&2.78&backg&def.&def.&kernib.h&def.&clean\\
			vim&8.1&fgui&def.&clean&bevl.c&def.&clean\\
			\bottomrule				
		\end{tabulary}	
		\begin{tablenotes}[para]
			
			\item[1] feature and file names have been shortned for brevity
		\end{tablenotes}
	\end{threeparttable}
	\vspace{-0.3cm}
\end{table}

\subsection{Results}
To make the predictions comparable, we mapped features to files and performed predictions for individual files and features in the test datasets, as explained in \secref{sec:rq3method}. We then filtered the results to analyze only defective files and features. \Tabref{tab:rq3Results} shows the filtered results. We expect that for each defective feature in a release, there must be at least one defective file. \Tabref{tab:rq3Results} shows that there are a total of 419 defective feature records (from all projects' test releases) mapped to a total of 628 defective file records. Note that the number 628 indicates total mappings between features and files and not unique files. For instance, the defective file \textit{kernel\_path\_branched.h
} in project \textit{blender}, contains implementation for 7 defective features in release 2.78 (\texttt{\_\_ao\_\_, \_\_background\_\_, \_\_denoising\_features\_\_, \_\_emission\_\_, \_\_kernel\_debug\_\_, \_\_shadow\_tricks\_\_, \_\_volume\_scatter\_\_}), and 5 defective features in release 2.79 (\texttt{\_\_ao\_\_, \_\_emission\_\_, \_\_holdout\_\_, \_\_subsurface\_\_, \_\_volume\_\_}), giving a total of 12 defective feature-file mapping records. For each mapping of features to files, we compare the proportion of features correctly predicted defective using feature-based defect prediction to the proportion of mapped defective files correctly predicted as defective using file-based defect prediction.

\insight{Our results show that out of the 419 defective feature records from all projects, feature-based defect prediction was able to correctly predict 274 (65\,\%) as defective, whereas none of the 628 defective file-records mapped to the features were correctly predicted.}

Since this result is extremely negative for file-based predictions, we examined the CSV file in which we recorded the predicted and actual labels for each file in the file-based test-dataset (available in our online appendix\,\cite{appendix:Online}). We found that there were 113 defective files in the dataset, of which only 4 (0.04\,\%) were correctly predicted as defective. These were: \texttt{interface\_eyedropper\_depth.c} (from blender, release 2.79), and \texttt{gimp-parallel.cc, gimpmeasuretool.c, performance-log-viewer.py} (from gimp, release 2\_10\_6). However, none of these four files had any feature mappings, hence, they do not appear in \tabref{tab:rq3Results}.

\looseness=-1
This result did not change despite trying different settings---balancing the file-based dataset, using the best classifier (RF), normalizing the data, and predicting with the best set of metrics---the top 75\,\% (see \secref{sec:rq2results}). \Tabref{tab:rq3ClassImbalance} shows that the file-based dataset is highly imbalanced, with defective files constituting only 2.4\,\% of all datatpoints in the dataset, whereas the feature-based dataset has a slightly higher percentage of defective instances (16.5\,\%).

Features are more abstract and cross-cutting entities whose implementation is often scattered across multiple files\,\cite{passos.ea:2018:tse}. Therefore, it is not surprising that the feature-based dataset has less imbalance than the file-based one, and thus has better performance. Our findings suggest that developers can obtain more precise defect prediction recommendations when applied at the granularity of features than files. Furthermore, when features are explicitly mapped to the code, as is our case, developers can be presented with files changed within a release (or commit) that are mapped to each feature predicted defective.
\setlength{\tymin}{3.5cm}
\begin{table}[tb]
	\centering
	\caption{Comparison of class-imbalance between feature-based and file-based datasets }
	\fontsize{8}{8}\selectfont
	
	\tablabel{tab:rq3ClassImbalance}
	\begin{threeparttable}
		\fontfamily{ptm}\selectfont
		
		\begin{tabulary}{\columnwidth}{@{}LRRRRR@{}}
			
			\toprule
			\theader{dataset}&\theader{clean}&\theader{\%}	&	\theader{defective}&\theader{\%}&\theader{total instances}	\\
			\midrule
			files-train&52,464  & 96.8\,\% &1,737 & 3.2\,\%&54,201 \\
			files-test&21,835  & 99.5\,\% &113 & 0.5\,\%&21,948 \\			
			files-total&74.299 & 97.6\,\% &1,850 & 2.4\,\%&76,149 \\
			\midrule
			features-train&8,606 &83.2\,\%&1,740&16.8\,\%&10,346\\
			features-test&2,403 &84.9\,\%&428&15.1\,\%&2,831\\
			features-total&11,009 &83.5\,\%&2,168&16.5\,\%&13,177\\
			
			\bottomrule				
		\end{tabulary}	
		\begin{tablenotes}[para]

		\end{tablenotes}
	\end{threeparttable}
	\vspace{-0.3cm}

\end{table}

\section{Support for Change- and Release-Based Defect Prediction (RQ4)}
\seclabel{sec:rq4ChangeBased}
\looseness=-1
We now discuss change- and release-based defect prediction, adressing RQ4 (\textit{To what extent can feature-based defect prediction support developers as they modify or release software features?}). While for RQ1, RQ2, and RQ3 we combined release-level data from all 12 subject projects, with RQ4, we investigate defect prediction for individual projects and individual releases and commits.

\subsection{Methodology}
\seclabel{sec:rq4method}
\looseness=-1
We generated defect prediction datasets at commit and release level. We used the two sets of datasets to understand the extent to which feature-based defect prediction can support developers as they \textit{modify} (during commits) and \textit{release} (during releases) features respectively. For commit-level datasets, we calculated the metrics in \tabref{tab:featureMetrics} at each commit based on prior commits. For instance, when generating data for the 5th commit in a project, the metric \metricName{fcomm}---distinct number of commits in which a feature has been modified---takes into account all past commits plus the current 5th commit.

Unlike previous research questions where we combined data from all projects and releases, and split them into train and test datasets, here, we generated training data per project, from all prior commits or releases (1 to nth), and predicted the defect proneness of features in subsequent commits or releases (n+1th) of the project. For instance, we trained Random Forest on the 1st commit, and predicted defect proneness of all features changed in the 2nd commit, then trained on data from the 1st and 2nd commit, and predicted for features in the 3rd commit, and so on. We did the same for releases. \Tabref{tab:rq4DatasetsPerProject} shows a summary of the number of commit-level and release-level datasets generated per project. Only project \textit{irssi} did not have test data for releases since it only had one release (1.0.6) from which we could extract defective features.

\setlength{\tymin}{3.5cm}
\begin{table}[b]
	\centering
	\caption{Number of commit and release datasets per project}
	\fontsize{8}{8}\selectfont
	\tablabel{tab:rq4DatasetsPerProject}
	\begin{threeparttable}
		\fontfamily{ptm}\selectfont
		
		\begin{tabulary}{\columnwidth}{@{}LRR@{}}
			
			\toprule
			\theader{project}&\theader{commit datasets}&\theader{release datasets}\\
			\midrule			
		
		blender&2,553&10\\
		busybox&438&13\\
		emacs&590&5\\
		gimp&441&13\\
		gnumeric&211&6\\
		gnuplot&4,082&4\\
		irssi&16&1\tnote{1}\\
		libxml2&127&9\\
		lighttpd&616&5\\
		mpsolve&98&3\\
		parrot&793&6\\
		vim&3,720&6\\
		\midrule
		Total&13,685&81\\
			\bottomrule				
		\end{tabulary}	
		\begin{tablenotes}[para]
			
			\item[1] irssi is missing results since it has 1 release, with clean features only
		\end{tablenotes}
	\end{threeparttable}
	\vspace{-0.3cm}
\end{table}
\subsection{Results}
\seclabel{sec:rq4results}
Using the commit and release level datasets we generated per project, we ran Random Forest on each test dataset and recorded resulting AUROC values to answer:
\begin{addmargin}[1em]{2em}
	\noindent \textit{RQ4.1: How effective is commit-level feature-based defect prediction for individual projects?} 
\end{addmargin}

\smallskip
\begin{addmargin}[1em]{2em}
	\noindent \textit{RQ4.2: How effective is release-level feature-based defect prediction for individual projects?} 
\end{addmargin}
\Figs~\fign{rq4CommitBOxPLot} and \fign{rq4ReleaseBOxPLot} show that commit-level defect prediction for all projects has a median AUROC score of 100\,\%, while release-level predictions have median AUROC over 95\,\% for all projects. Furthermore, \figref{rq4TrendsOfCommitLeveklPredictions} shows the trends of the commit-level predictions for 8 projects (we only show 8 here, the other 3 projects had similar trends). For most projects, prediction performance sharply drops when several new files are introduced, especially in the beginning of the project, however, such drops occur for few commits. 

\insight{Our results suggest that feature-based defect prediction can support developers with recommendations for defect-prone features, with AUROC over 95\,\%, at both commit and release-level. We also observe that the AUROC for release-level prediction became better when considering individual projects instead of combining training and test data from all projects as was done for RQ1.}

\begin{figure}[b]
	\centering
	\includegraphics[width=\columnwidth]{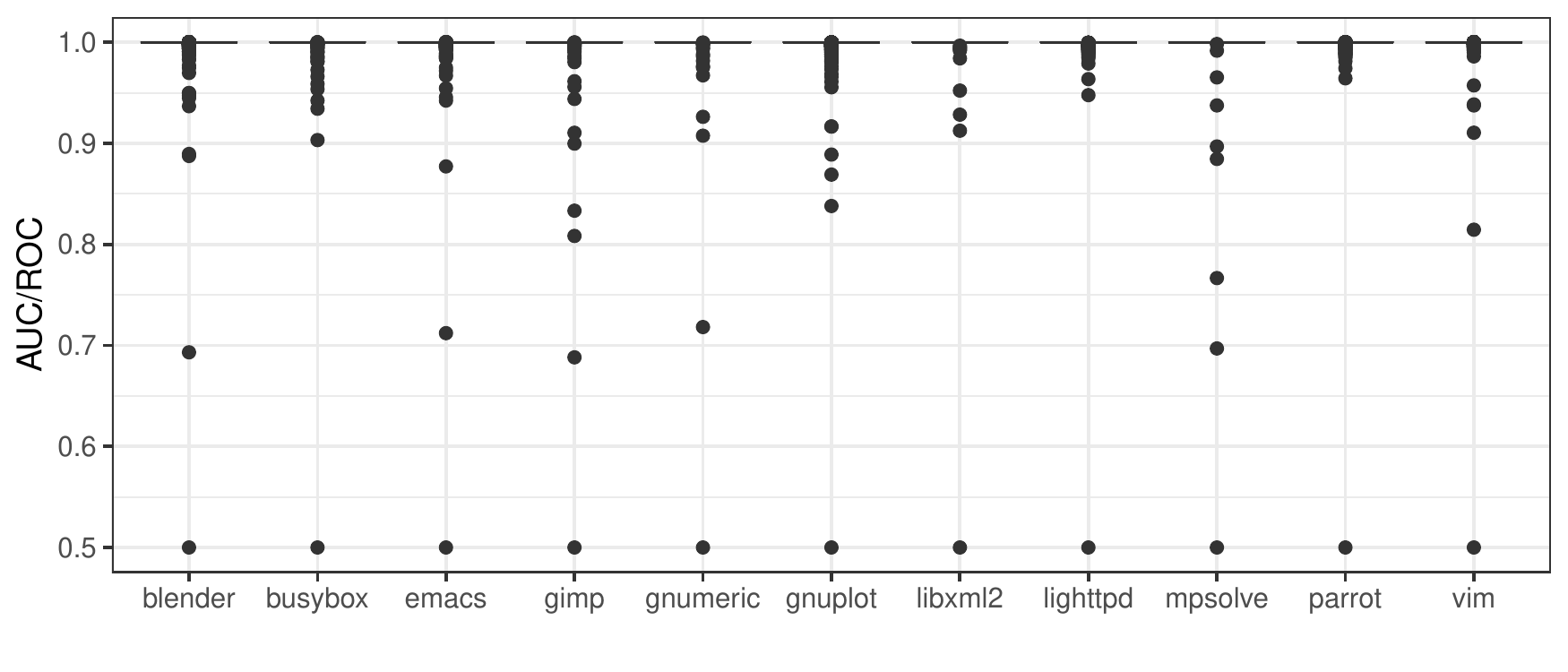}
	\vspace{-.7cm}
	\caption{Commit-level prediction per project}
	\label{fig:rq4CommitBOxPLot}
	\vspace{-.4cm}
\end{figure}
\begin{figure}[b]
	\centering
	\includegraphics[width=\columnwidth]{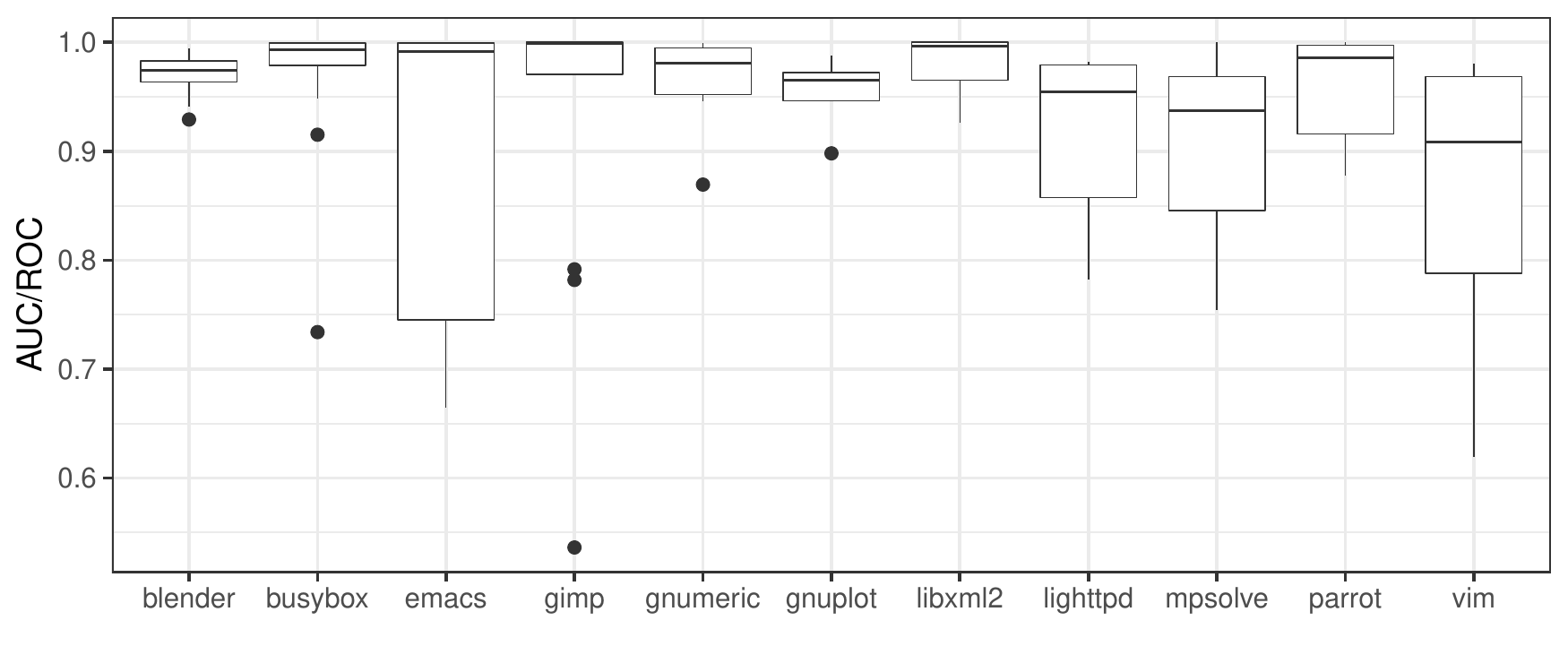}
	\vspace{-.7cm}
	\caption{Release-level prediction per project}
	\label{fig:rq4ReleaseBOxPLot}
	\vspace{-.4cm}
\end{figure}
\begin{figure}[t]
	\centering
	\includegraphics[width=\columnwidth]{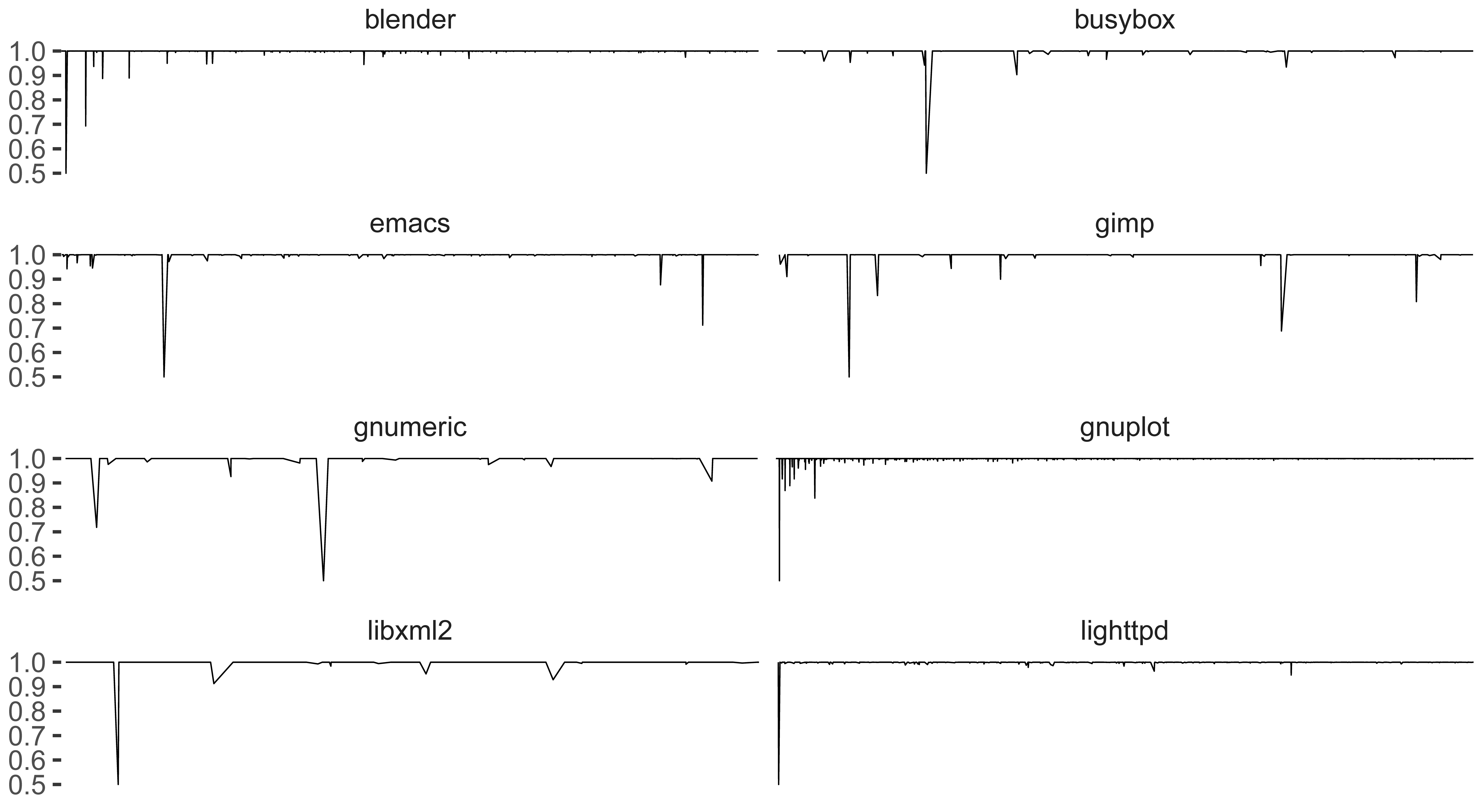}
	\vspace{-.7cm}
	\caption{Trends of commit-level prediction in 8 of the 12 projects}
	\label{fig:rq4TrendsOfCommitLeveklPredictions}
	\vspace{-.4cm}
\end{figure}

\section{Cross-Project Defect Prediction (RQ5)}
\seclabel{rq5Crossproject}
In this section, we address cross-project defect prediction. We discuss and answer RQ5: \textit{To what extent can feature-based defect prediction models be reused across projects without re-training?}  
\subsection{Methodology}
\seclabel{rq5method}
To address RQ5, we performed the following steps.
\setlength{\tymin}{3.5cm}
\begin{table}[b]
	\centering
	\caption{Project combinations for cross-project model reuse}
	\fontsize{8}{8}\selectfont
	\tablabel{tab:rq5ProjectCombinations}
	\begin{threeparttable}
		\fontfamily{ptm}\selectfont
		
		\begin{tabulary}{\columnwidth}{@{}CCR@{}}
			
			\toprule
		\theader{train on (\# of projects)}& \theader{predict for (\# of projects)}&\theader{combinations}\\
		\midrule
		1& 11&11\\
		2& 10&127\\
		3& 9&650\\
		4& 8&1,970\\
		5& 7&3,955\\
		6&6 &5,543\\
		7& 4&5,544\\
		8& 4&3,960\\
		9& 3&1,980\\
		10& 2&660\\
		11& 1 &132\\
		\midrule
		&total&24,532\\
		
			\bottomrule				
		\end{tabulary}	
		\begin{tablenotes}[para]
			
		\end{tablenotes}
	\end{threeparttable}
	\vspace{-0.3cm}
\end{table}

\begin{itemize}
	\item [1] For each of our 12 subject projects, we created two datasets, one with release-level data for all releases in the project and the other with commit-level data for all commits in the project. We based this on the datasets created when addressing RQ4 (support for change-based predictions). However, unlike the 13,685 commit-level and 81 release-level datasets used in RQ4, which were generated per project, here, we combine data from commits or releases in a project to generate one dataset respectively. In total, we created 24 datasets (12 with commit-level data per project, and 12 with release-level data per project).
	\item[2] Using the list of our 12 subject projects, we generated all possible combinations of $x:y$, where $x$ is the number of projects providing training data used to predict for each of the remaining $y$ projects. Both $x$ and $y$ range from 1 to 11. For instance, for the ratio 1:11 (i.e., train on 1 project and predict for each of the remaining 11 projects), we generated all combinations where each project provided training data used to predict for the other 11 projects. Similarly, for the ratio 4:8, we generated all combinations of 4 projects used to provide training data while predicting for the remaining 8. In any case, the projects in the training set are always excluded from the test set. \Tabref{tab:rq5ProjectCombinations} shows all possible combinations of projects we generated for each ratio. In total, we generated 24,532 training combinations  and 24,532 test combinations.
	\item[3] For each training set combination, we created the corresponding ARFF dataset file using data from all projects in the set; one at commit level and another at release-level. For instance if a training set combination consists of projects \textit{blender, emacs,} and \textit{busybox}, we create commit-level and release-level training datasets with data from these three projects. Therefore, we created 24,532 commit-level training datasets and 24,532 release-level training datasets; totaling 49,064 datasets.
	\item[4] Finally, we use each training dataset to predict for each project in the test set individually. We run the predictions on the project datasets generated in step 1 and record the \auroc values obtained. For instance, for training set combination \textit{blender, emacs, busybox}, we would run predictions on each dataset for the remaining projects (\textit{gimp, gnumeric, gnuplot, irssi, lib2xml, lighthtpd, parrot, vim}).
\end{itemize}
\subsection{Results}
\seclabel{rq5results}
\begin{figure}[b]
	\centering
	\includegraphics[width=\columnwidth]{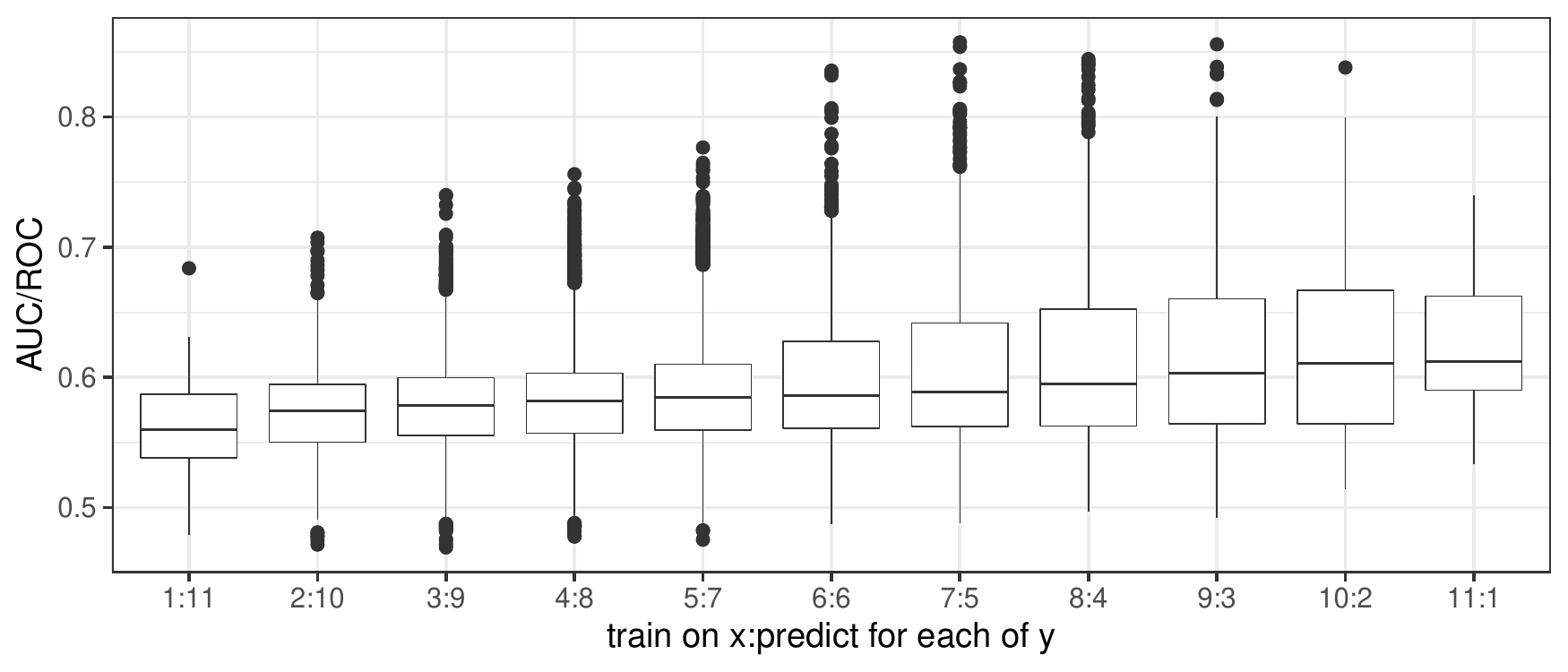}
	\vspace{-.7cm}
	\caption{Commit-level cross-project defect prediction per test project}
	\label{fig:rq5CommitBOxPLot}
	\vspace{-.4cm}
\end{figure}

\begin{figure}[t]
	\centering
	\includegraphics[width=\columnwidth]{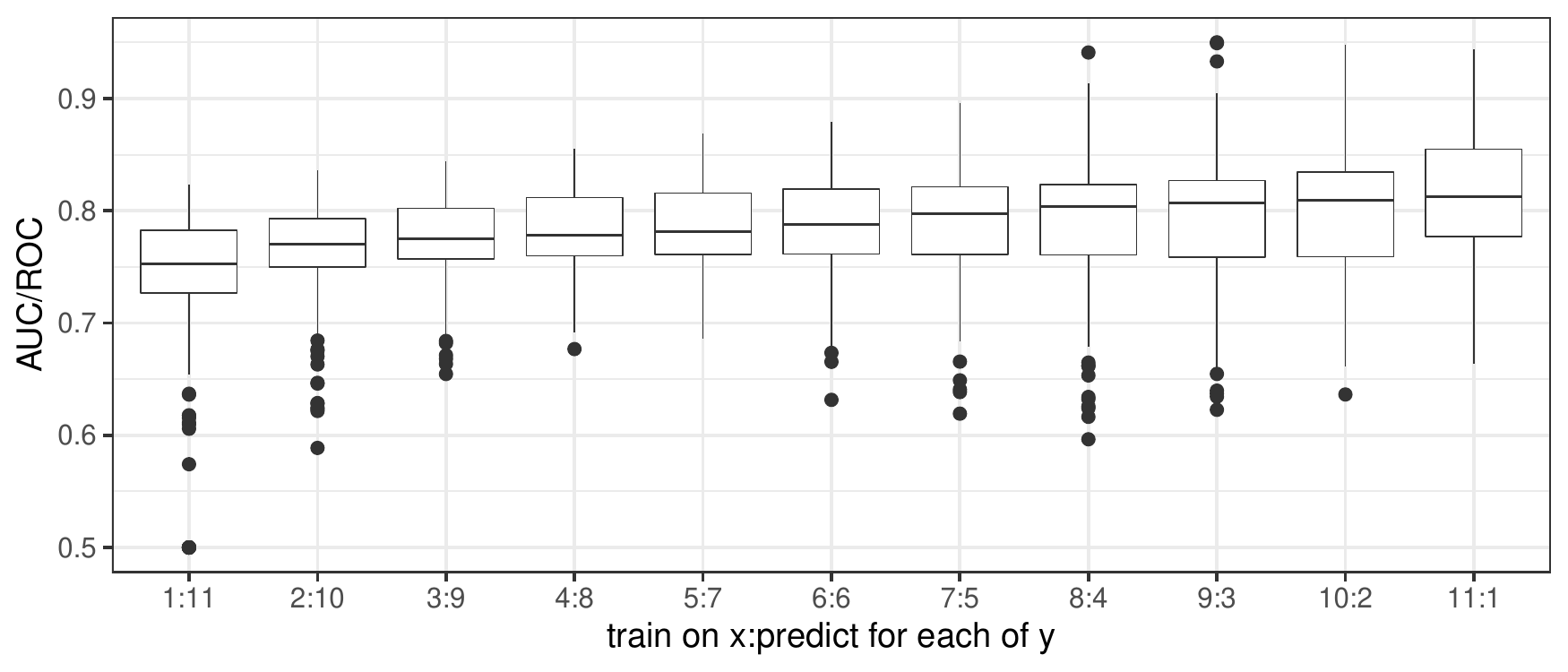}
	\vspace{-.7cm}
	\caption{Release-level cross-project defect prediction per test project}
	\label{fig:rq5ReleaseBOxPLot}
	\vspace{-.4cm}
\end{figure}
\Figs~\fign{rq5CommitBOxPLot} and \fign{rq5ReleaseBOxPLot} show that release-level cross-project predictions outperformed commit-level predictions; with the former scoring  median \auroc ranging from 0.75 (ratio 1:11) to 0.81 (ratio 11:1), while the latter scored median \auroc ranging from 0.56 (ratio 1:11) to 0.61 (ratio 11:1). As expected, the higher the number of projects in the training set, the better the prediction performance.

For the release-level predictions shown in \figref{rq5ReleaseBOxPLot}, we observe no marked differences in median \auroc values when training with 2, 3, 4, 5, or 6 projects. Similarly, there are no observable differences between median \auroc values obtained when training with 7,8,9, or 10 projects.

\insight{Our results suggest that release-level feature-based defect prediction models can be reused across projects even when trained on single projects, and can deliver median performance values ranging from 75\,\% to 81\,\%.}

We further analyzed all combinations of the ratio 1:11, given the relatively good median \auroc value we obtained (0.75). Such a high median score indicates the potential to reuse feature-based defect prediction models trained on one representative project only. 

\begin{figure}[ht]
	\centering
	\includegraphics[width=\columnwidth]{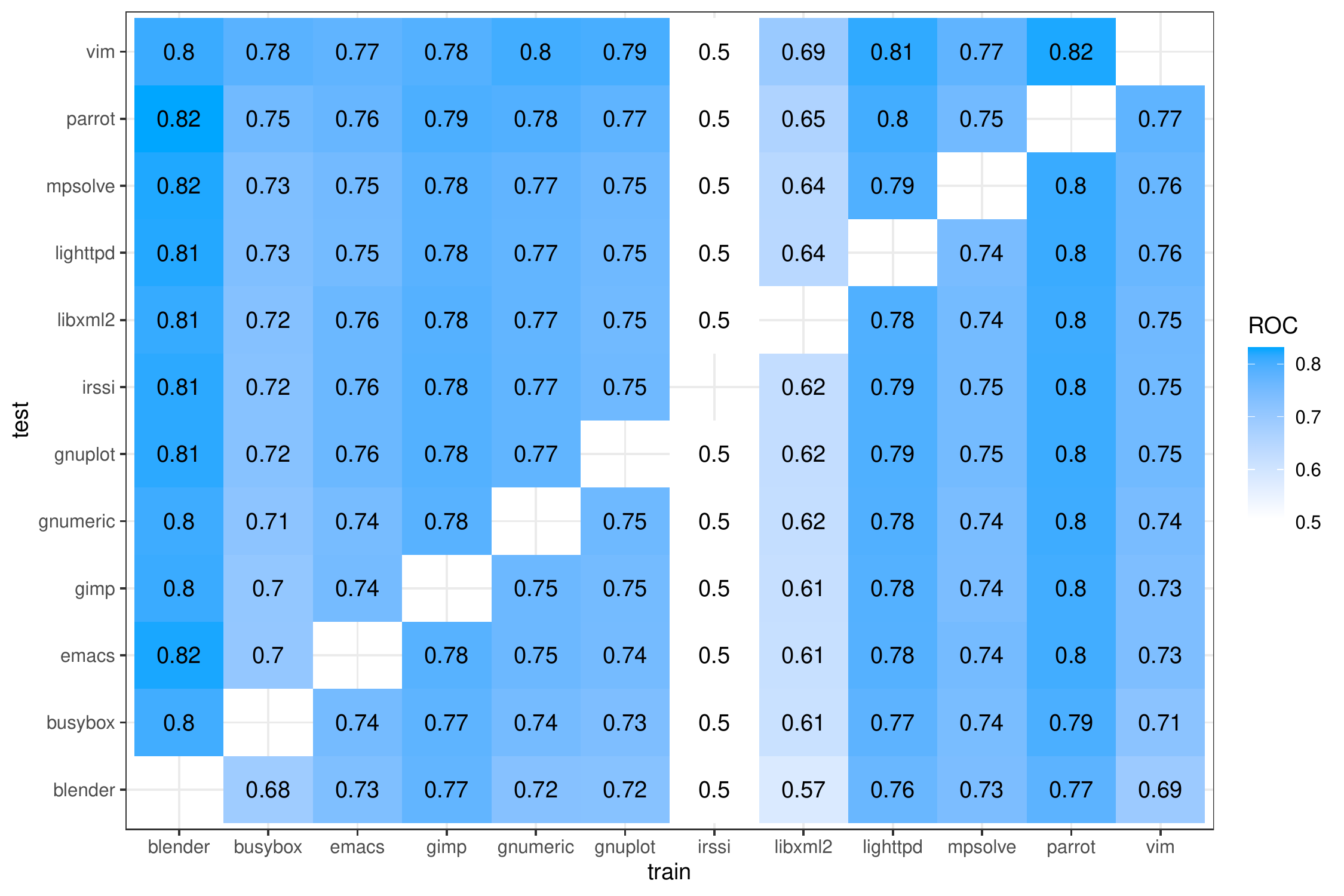}
	\vspace{-.7cm}
	\caption{Release-level cross-project prediction using training data from one project only}
	\label{fig:rq5Release111HeatMap}
	\vspace{-.4cm}
\end{figure}

\Figref{rq5Release111HeatMap} shows a heatmap of all \auroc values obtained for all combinations of training on 1 project and predicting for the remaining 11 (1:11). We observed that some project were more useful for predicting defects in other projects. Particularly, \textit{blender} and \textit{parrot} where the top 2 training data sources, followed by \textit{lighthttpd} and \textit{gimp} respectively. Interestingly, with \textit{blender} alone, we were able to predict for all other projects with \auroc ranging from 80\,\% to 82\,\%.

\setlength{\tymin}{3.5cm}
\begin{table}[tb]
	\centering
	\caption{Total number of release-level datapoints per project}
	\fontsize{8}{8}\selectfont
	\tablabel{tab:rq5DatapointsPerProject}
	\begin{threeparttable}
		\fontfamily{ptm}\selectfont
		
		\begin{tabulary}{\columnwidth}{@{}LR@{}}
			
			\toprule
		\theader{project}&\theader{instances}\\
		\midrule
		blender&5,245\\
		busybox&876\\
		emacs&508\\
		gimp&378\\
		gnumeric&694\\
		gnuplot&1,112\\
		irssi&8\\
		libxml2&217\\
		lighttpd&637\\
		mpsolve&70\\
		parrot&693\\
		vim&2,739\\
			\bottomrule				
		\end{tabulary}	
		\begin{tablenotes}[para]

		\end{tablenotes}
	\end{threeparttable}
	\vspace{-0.3cm}
\end{table}
\Tabref{tab:rq5DatapointsPerProject} shows the total number of release-level data points for each project. We find that, despite \textit{parrot} and \textit{lighttpd} having fewer datapoints than other projects, such as \textit{vim} and \textit{gnuplot}, they still performed better; thus indicating that other project characteristics contribute to better  cross-project predictions in addition to the number of training datapoints (c.f. see \textit{irssi}'s poor performance in \figref{rq5Release111HeatMap}; has only 8 datapoints). Therefore, an investigation of project properties relevant for cross project prediction would be worthwhile future work.

\section{Threats to Validity}

\looseness=-1
\parhead{External validity}
To mitigate overfitting of our models, a key threat in machine learning, we used the typical separation of the dataset into test data and training data\,\cite{domingos2012few}.
Another standard technique to mitigate overfitting, cross validation, is not applicable to scenario, since it leads to the problematic situation of ``using the future to predict the past,'' which is unrealistic for practical applications\,\cite{jimenez2019importance}.
While providing improved techniques for avoiding overfitting in defect prediction contexts is an open research issue, we observe good predictive ability for 12 systems of diverse context and size, which gives us some confidence that our models are not severely affected by overfitting.

Despite diversity and size of the projects in our dataset, studying a broader selection of software projects is desirable, as it would increase the generalizability of our findings.
We intend to aggregate larger datasets in future work, which would also contribute to our ongoing community initiative towards more mature benchmarks for techniques in the context of evolving variant-rich systems\,\cite{SMK19}.

\parhead{Internal validity} 
As observed in \secref{sec:dataset}, some of our automatically retrieved features were not meaningful, as they  represented "header features", in the style of a certain C pattern.
While we manually processed all identified features to remove header features, it is possible that in some projects header features are not explicitly identified by names.
A possible solution could be enabled by a tool that automatically analyses the code to detect header features. Such a tool does not exist at the moment.
Still, the manual removal of the recognizable header features allowed us to  reduce the amount of these noise datapoints. Furthermore, we take a conservative approach by considering only features referenced through \hashifdef and \ifndef. We also do not explicitly exclude standard predefined macros such as \texttt{\small{\_\_FILE\_\_}}, \texttt{\small{\_\_LINE\_\_}}, etc., since we treat them as features with associated code if they are referenced through our selected preprocessor macros above.

\looseness=-1
During dataset creation, we relied on a mapping from all features changed in a particular release to the associated files.
This mapping is obtained from analyzing all commit change sets within the release.
Thus, a feature is considered relevant if it is mentioned in a diff (either within a changed line or in the context provided with change lines, which, per default, extends to three lines before and after changed lines).
While this heuristic may exclude preprocessor macros outside the provided context, it has less impact on our results since it only potentially reduces the number of learning examples (features). Still, extending the implementation to take into account all features in changed files is subject to future work.

\parhead{Construct validity} 
Our ground truth for the identification of defective and clean features relies on an available heuristic technique, the SZZ algorithm. 
An associated threat is concerned with possible imprecisions of this algorithm.
According to a recent study\,\cite{Wen2019},  available implementations of SZZ, including those of PyDriller, can identify only about 69\% of all bug-introducing commits. In addition, about 64\% of the identified commits were found to be incorrectly identified.
These imprecisions arise from violations to implicit assumptions of the SZZ algorithm.
Furthermore, the authors of the study empirically found that the results of eight out of ten earlier studies were significantly influenced by the imprecise algorithm\,\cite{Wen2019}. This may, therefore, also apply to this work. However, there is currently no alternative method for identifying bug-introducing commits. Whenever an improved method becomes available, we will repeat the main steps of this work, taking the new method into account, and compare with our results.

\section{Conclusion}

\looseness=-1
We presented a systematic investigation of feature-based defect prediction.
To predict possible software defects on the granularity of both features and files, we constructed a dataset based on 12 real revision histories of feature-based software projects.
We systematically investigated feature engineering and finally derived two new carefully crafted metric sets, one solely based on process metrics, one based on a combination of process and structure metrics.
We evaluated the predictive ability of seven classifiers in combination with our new metric sets and an additional metric set from a previous work. We also evaluated different scenarios: whether our feature-based metrics improve file-based predictions, how feature-based predictions compare to file-based ones, to what extent feature-based defect prediction can be performed per commit or release of a single project, and to what extent prediction models can be reused across projects without retraining. We conclude:

\begin{itemize}[leftmargin=*]
	\item Using a more diverse metrics set, comprising structure and process metrics, leads to more robustness and on-average better prediction results.
	\item Simple classifiers, such as NB, in combination with a simpler metric set (process metrics only) can produce high-quality results; however, at the cost of robustness.
	\item Enabled by our most advanced metric set, we find two best-performing models (precision, recall, robustness): one based on a random forest classifier, the other based on a neutral network. 
	\item Even though our feature-based metrics improve file-based predictions by a non-significant margin, we found them to have a dominant effect on the predictions compared to file-based metrics.
	\item Based on the mapping between features and files, feature-based defect prediction correctly yields a higher number of defective files than file-based prediction.
	\item It is possible to perform defect prediction at commit and release-levels for individual projects with \auroc of over 95\,\%
	\item Feature-based defect prediction models, trained on release-level data, can be reused across projects, with median \auroc over 75\,\% without retraining, even when trained from a single representative project.
\end{itemize}

\looseness=-1
In the future, we would like to investigate what project characteristics make some projects better sources of training data than others. Such a study would be useful in formulating general guidelines for practitioners who may want to predict defects using models trained on existing projects with historic data.
Another direction would be to consider predicting unwanted feature interactions\,\cite{apel2014feature}, which are special kinds of bugs that, when taken into account, may improve the predictive ability of defect prediction techniques, as well as provide meaningful insights regarding whether some machine learning classifiers perform better than others on specific kinds of bugs. Here, we could apply techniques for identifying variability-aware bugs\,\cite{abal201442}, or 
automatically generate test cases for features and use them as partial specifications\,\cite{leuson2020testconflicts} to identify unwanted feature behavior.

\begingroup
\sloppy

\balance

\bibliographystyle{IEEEtran}

\normalsize
\looseness=-1
\bibliography{main}

\begin{thebibliography}{10}
\providecommand{\url}[1]{#1}
\csname url@samestyle\endcsname
\providecommand{\newblock}{\relax}
\providecommand{\bibinfo}[2]{#2}
\providecommand{\BIBentrySTDinterwordspacing}{\spaceskip=0pt\relax}
\providecommand{\BIBentryALTinterwordstretchfactor}{4}
\providecommand{\BIBentryALTinterwordspacing}{\spaceskip=\fontdimen2\font plus
\BIBentryALTinterwordstretchfactor\fontdimen3\font minus
  \fontdimen4\font\relax}
\providecommand{\BIBforeignlanguage}[2]{{%
\expandafter\ifx\csname l@#1\endcsname\relax
\typeout{** WARNING: IEEEtran.bst: No hyphenation pattern has been}%
\typeout{** loaded for the language `#1'. Using the pattern for}%
\typeout{** the default language instead.}%
\else
\language=\csname l@#1\endcsname
\fi
#2}}
\providecommand{\BIBdecl}{\relax}
\BIBdecl

\bibitem{Challagulla2008}
V.~U.~B. Challagulla, F.~B. Bastani, I.~L. Yen, and R.~A. Paul, ``{Empirical
  assessment of machine learning based software defect prediction
  techniques},'' \emph{IJAIT}, vol.~17, no.~2, pp. 389--400, 2008.

\bibitem{Alsaeedi2019}
A.~Alsaeedi and M.~Z. Khan, ``Software defect prediction using supervised
  machine learning and ensemble techniques: A comparative study,'' \emph{JSEA},
  vol.~12, no.~05, pp. 85--100, 2019.

\bibitem{Hammouri2018}
A.~Hammouri, M.~Hammad, M.~Alnabhan, and F.~Alsarayrah, ``Software bug
  prediction using machine learning approach,'' \emph{IJACSA}, vol.~9, no.~2,
  2018.

\bibitem{Son2019}
L.~Son, N.~Pritam, M.~Khari, R.~Kumar, P.~Phuong, and P.~Thong, ``Empirical
  study of software defect prediction: A systematic mapping,'' \emph{Symmetry},
  vol.~11, no.~2, p. 212, Feb. 2019.

\bibitem{Apel2013}
S.~Apel, D.~Batory, C.~K{\"{a}}stner, and G.~Saake, \emph{{Feature-Oriented
  Software Product Lines}}.\hskip 1em plus 0.5em minus 0.4em\relax Springer,
  2013.

\bibitem{berger2015feature}
T.~Berger, D.~Lettner, J.~Rubin, P.~Gr{\"u}nbacher, A.~Silva, M.~Becker,
  M.~Chechik, and K.~Czarnecki, ``What is a feature? a qualitative study of
  features in industrial software product lines,'' in \emph{SPLC}, 2015.

\bibitem{kang.ea:1990:foda}
K.~C. Kang, S.~Cohen, J.~Hess, W.~Nowak, and S.~Peterson, ``{Feature-Oriented
  Domain Analysis {(FODA)} Feasibility Study},'' Carnegie-Mellon University,
  Tech. Rep. CMU/SEI-90-TR-21, 1990.

\bibitem{damir2019principles}
D.~Nesic, J.~Krueger, S.~Stanciulescu, and T.~Berger, ``Principles of feature
  modeling,'' in \emph{FSE}, 2019.

\bibitem{larman2008scaling}
C.~Larman, \emph{Scaling lean \& agile development: thinking and organizational
  tools for large-scale Scrum}.\hskip 1em plus 0.5em minus 0.4em\relax Pearson
  Education India, 2008.

\bibitem{passos.ea:2018:tse}
L.~Passos, R.~Queiroz, M.~Mukelabai, T.~Berger, S.~Apel, K.~Czarnecki, and
  J.~Padilla, ``A study of feature scattering in the linux kernel,''
  \emph{TSE}, vol.~47, no.~01, pp. 146--164, 2018.

\bibitem{bruns2005foundations}
G.~Bruns, ``Foundations for features,'' in \emph{Feature Interactions in
  Telecommunications and Software System}.\hskip 1em plus 0.5em minus
  0.4em\relax IOS Press, 2005, pp. 3--11.

\bibitem{zave:2004:features}
P.~Zave, ``{FAQ Sheet on Feature Interactions},'' Available at
  \url{http://www.research.att.com/~pamela/faq.html}, 2004.

\bibitem{apel2014feature}
S.~Apel, J.~M. Atlee, L.~Baresi, and P.~Zave, ``Feature interactions: the next
  generation (dagstuhl seminar 14281),'' in \emph{Dagstuhl Reports}, vol.~4,
  no.~7.\hskip 1em plus 0.5em minus 0.4em\relax Schloss
  Dagstuhl-Leibniz-Zentrum fuer Informatik, 2014.

\bibitem{Moser2008}
R.~Moser, W.~Pedrycz, and G.~Succi, ``A comparative analysis of the efficiency
  of change metrics and static code attributes for defect prediction,'' in
  \emph{ICSE}, 2008.

\bibitem{kamei2016studying}
Y.~Kamei, T.~Fukushima, S.~McIntosh, K.~Yamashita, N.~Ubayashi, and A.~E.
  Hassan, ``Studying just-in-time defect prediction using cross-project
  models,'' \emph{EMSE}, vol.~21, no.~5, pp. 2072--2106, 2016.

\bibitem{zimmermann2009cross}
T.~Zimmermann, N.~Nagappan, H.~Gall, E.~Giger, and B.~Murphy, ``Cross-project
  defect prediction: a large scale experiment on data vs. domain vs. process,''
  in \emph{ESEC/FSE}, 2009.

\bibitem{appendix:Online}
{The Authors}, ``{Online Appendix},''
  \url{https://bitbucket.org/easelab/onlineappendixdefectpred}, 2020.

\bibitem{struder2020feature}
S.~Str{\"u}der, M.~Mukelabai, D.~Str{\"u}ber, and T.~Berger, ``Feature-oriented
  defect prediction,'' in \emph{SPLC}, 2020.

\bibitem{nam2013transfer}
J.~Nam, S.~J. Pan, and S.~Kim, ``Transfer defect learning,'' in \emph{ICSE},
  2013.

\bibitem{rahman2011bugcache}
F.~Rahman, D.~Posnett, A.~Hindle, E.~Barr, and P.~Devanbu, ``Bugcache for
  inspections: hit or miss?'' in \emph{ESEC/FSE}, 2011.

\bibitem{d2010extensive}
M.~D'Ambros, M.~Lanza, and R.~Robbes, ``An extensive comparison of bug
  prediction approaches,'' in \emph{MSR}, 2010.

\bibitem{zimmermann2008predicting}
T.~Zimmermann and N.~Nagappan, ``Predicting defects using network analysis on
  dependency graphs,'' in \emph{ICSE}, 2008.

\bibitem{menzies2006data}
T.~Menzies, J.~Greenwald, and A.~Frank, ``Data mining static code attributes to
  learn defect predictors,'' \emph{TSE}, vol.~33, no.~1, pp. 2--13, 2006.

\bibitem{akiyama1971example}
F.~Akiyama, ``An example of software system debugging.'' in \emph{IFIP Congress
  (1)}, vol.~71, 1971, pp. 353--359.

\bibitem{mccabe1976complexity}
T.~J. McCabe, ``A complexity measure,'' \emph{TSE}, no.~4, pp. 308--320, 1976.

\bibitem{halstead1977elements}
M.~H. Halstead \emph{et~al.}, \emph{Elements of software science}.\hskip 1em
  plus 0.5em minus 0.4em\relax Elsevier New York, 1977, vol.~7.

\bibitem{d2012evaluating}
M.~D'Ambros, M.~Lanza, and R.~Robbes, ``Evaluating defect prediction
  approaches: a benchmark and an extensive comparison,'' \emph{EMSE}, vol.~17,
  no. 4-5, pp. 531--577, 2012.

\bibitem{bird2011don}
C.~Bird, N.~Nagappan, B.~Murphy, H.~Gall, and P.~Devanbu, ``Don't touch my
  code! examining the effects of ownership on software quality,'' in
  \emph{ESEC/FSE}, 2011.

\bibitem{lee2011micro}
T.~Lee, J.~Nam, D.~Han, S.~Kim, and H.~P. In, ``Micro interaction metrics for
  defect prediction,'' in \emph{ESEC/FSE}, 2011.

\bibitem{bacchelli2010popular}
A.~Bacchelli, M.~D'Ambros, and M.~Lanza, ``Are popular classes more defect
  prone?'' in \emph{FASE}, 2010.

\bibitem{hassan2009predicting}
A.~E. Hassan, ``Predicting faults using the complexity of code changes,'' in
  \emph{ICSE}, 2009.

\bibitem{nagappan2005use}
N.~Nagappan and T.~Ball, ``Use of relative code churn measures to predict
  system defect density,'' in \emph{ICSE}, 2005.

\bibitem{shivaji2012reducing}
S.~Shivaji, E.~J. Whitehead, R.~Akella, and S.~Kim, ``Reducing features to
  improve code change-based bug prediction,'' \emph{TSE}, vol.~39, no.~4, pp.
  552--569, 2012.

\bibitem{Rahman2013}
F.~Rahman and P.~Devanbu, ``How, and why, process metrics are better,'' in
  \emph{ICSE}.\hskip 1em plus 0.5em minus 0.4em\relax {IEEE}, May 2013, pp.
  432--441.

\bibitem{lessmann2008benchmarking}
S.~Lessmann, B.~Baesens, C.~Mues, and S.~Pietsch, ``Benchmarking classification
  models for software defect prediction: A proposed framework and novel
  findings,'' \emph{TSE}, vol.~34, no.~4, pp. 485--496, 2008.

\bibitem{mende2010replication}
T.~Mende, ``Replication of defect prediction studies: problems, pitfalls and
  recommendations,'' in \emph{PROMISE}, 2010.

\bibitem{arisholm2007data}
E.~Arisholm, L.~C. Briand, and M.~Fuglerud, ``Data mining techniques for
  building fault-proneness models in telecom java software,'' in \emph{ISSRE},
  2007.

\bibitem{kamei2012large}
Y.~Kamei, E.~Shihab, B.~Adams, A.~E. Hassan, A.~Mockus, A.~Sinha, and
  N.~Ubayashi, ``A large-scale empirical study of just-in-time quality
  assurance,'' \emph{TSE}, vol.~39, no.~6, pp. 757--773, 2012.

\bibitem{kim2008classifying}
S.~Kim, E.~J. Whitehead~Jr, and Y.~Zhang, ``Classifying software changes: Clean
  or buggy?'' \emph{TSE}, vol.~34, no.~2, pp. 181--196, 2008.

\bibitem{he2012investigation}
Z.~He, F.~Shu, Y.~Yang, M.~Li, and Q.~Wang, ``An investigation on the
  feasibility of cross-project defect prediction,'' \emph{J. ASE}, vol.~19,
  no.~2, pp. 167--199, 2012.

\bibitem{klas2010transparent}
M.~Kl{\"a}s, F.~Elberzhager, J.~M{\"u}nch, K.~Hartjes, and O.~von Graevemeyer,
  ``Transparent combination of expert and measurement data for defect
  prediction: an industrial case study,'' in \emph{ICSE}, 2010, pp. 119--128.

\bibitem{fenton2008effectiveness}
N.~Fenton, M.~Neil, W.~Marsh, P.~Hearty, {\L}.~Radli{\'n}ski, and P.~Krause,
  ``On the effectiveness of early life cycle defect prediction with bayesian
  nets,'' \emph{EMSE}, vol.~13, no.~5, p. 499, 2008.

\bibitem{marcus2008using}
A.~Marcus, D.~Poshyvanyk, and R.~Ferenc, ``Using the conceptual cohesion of
  classes for fault prediction in object-oriented systems,'' \emph{TSE},
  vol.~34, no.~2, pp. 287--300, 2008.

\bibitem{ostrand2005predicting}
T.~J. Ostrand, E.~J. Weyuker, and R.~M. Bell, ``Predicting the location and
  number of faults in large software systems,'' \emph{TSE}, vol.~31, no.~4, pp.
  340--355, 2005.

\bibitem{hata2012bug}
H.~Hata, O.~Mizuno, and T.~Kikuno, ``Bug prediction based on fine-grained
  module histories,'' in \emph{ICSE}, 2012.

\bibitem{giger2012method}
E.~Giger, M.~D'Ambros, M.~Pinzger, and H.~C. Gall, ``Method-level bug
  prediction,'' in \emph{EASE}.\hskip 1em plus 0.5em minus 0.4em\relax IEEE,
  2012, pp. 171--180.

\bibitem{Queiroz2016}
R.~Queiroz, T.~Berger, and K.~Czarnecki, ``Towards predicting feature defects
  in software product lines,'' in \emph{FOSD}, 2016.

\bibitem{apel:2013:fospl}
S.~Apel, D.~Batory, C.~K{\"a}stner, and G.~Saake, \emph{Feature-Oriented
  Software Product Lines}.\hskip 1em plus 0.5em minus 0.4em\relax Springer,
  2013.

\bibitem{pereira2019learning}
J.~A. Pereira, H.~Martin, M.~Acher, J.-M. J{\'e}z{\'e}quel, G.~Botterweck, and
  A.~Ventresque, ``Learning software configuration spaces: A systematic
  literature review,'' \emph{arXiv preprint arXiv:1906.03018}, 2019.

\bibitem{siegmund:splc:2011}
N.~Siegmund, M.~Rosenmuller, C.~K\"{a}stner, P.~G. Giarrusso, S.~Apel, and
  S.~S. Kolesnikov, ``Scalable prediction of non-functional properties in
  software product lines,'' in \emph{SPLC}, 2011.

\bibitem{Temple2016}
P.~Temple, J.~A. Galindo, M.~Acher, and J.~J{\'{e}}z{\'{e}}quel, ``Using
  machine learning to infer constraints for product lines,'' in \emph{SPLC},
  2016.

\bibitem{nadi.ea:2015:tse}
S.~Nadi, T.~Berger, C.~K{\"a}stner, and K.~Czarnecki, ``Where do configuration
  constraints stem from? an extraction approach and an empirical study,''
  \emph{TSE}, vol.~41, no.~8, pp. 820--841, 2015.

\bibitem{temple2019towards}
P.~Temple, M.~Acher, G.~Perrouin, B.~Biggio, J.-M. J{\'e}z{\'e}quel, and
  F.~Roli, ``Towards quality assurance of software product lines with
  adversarial configurations,'' in \emph{SPLC}, 2019.

\bibitem{ghofrani2019applying}
J.~Ghofrani, E.~Kozegar, A.~L. Fehlhaber, and M.~D. Soorati, ``Applying product
  line engineering concepts to deep neural networks,'' in \emph{SPLC}, 2019.

\bibitem{ghofrani2019reusability}
J.~Ghofrani, E.~Kozegar, A.~Bozorgmehr, and M.~D. Soorati, ``Reusability in
  artificial neural networks: an empirical study,'' in \emph{SPLC}, 2019.

\bibitem{el2019metrics}
S.~El-Sharkawy, N.~Yamagishi-Eichler, and K.~Schmid, ``Metrics for analyzing
  variability and its implementation in software product lines: A systematic
  literature review,'' \emph{IST}, vol. 106, pp. 1--30, 2019.

\bibitem{passos.ea:2013:evolution}
L.~Passos, K.~Czarnecki, S.~Apel, A.~W\k{a}sowski, C.~K\"{a}stner, and J.~Guo,
  ``{Feature-Oriented Software Evolution},'' in \emph{VAMOS}, 2013.

\bibitem{Liebig2010}
J.~Liebig, S.~Apel, C.~Lengauer, C.~K{\"a}stner, and M.~Schulze, ``An analysis
  of the variability in forty preprocessor-based software product lines,'' in
  \emph{ICSE}, 2010.

\bibitem{Berger:2014:TSA:2556624.2556641}
T.~Berger and J.~Guo, ``Towards system analysis with variability model
  metrics,'' in \emph{VaMoS}, 2013.

\bibitem{strueber2020modelsvar}
D.~Strueber, A.~Anjorin, and T.~Berger, ``Variability representations in class
  models: An empirical assessment,'' in \emph{MODELS}, 2020.

\bibitem{Medeiros:2015wg}
F.~Medeiros, C.~K{\"a}stner, M.~Ribeiro, S.~Nadi, and R.~Gheyi, ``{The
  Love/Hate Relationship with the C Preprocessor - An Interview Study},'' in
  \emph{ECOOP}, 2015, pp. 495--518.

\bibitem{Batory:2004bw}
D.~Batory, J.~N. Sarvela, and A.~Rauschmayer, ``{Scaling Step-Wise
  Refinement},'' \emph{TSE}, vol.~30, no.~6, pp. 355--371, 2004.

\bibitem{mukelabai.ea:2018:analysis}
M.~Mukelabai, D.~Ne{\v s}i{\'c}, S.~Maro, T.~Berger, and J.-P. Stegh\"ofer,
  ``Tackling combinatorial explosion: A study of industrial needs and practices
  for analyzing highly configurable systems,'' in \emph{ASE}, 2018.

\bibitem{hunsen2016preprocessor}
C.~Hunsen, B.~Zhang, J.~Siegmund, C.~K{\"a}stner, O.~Le{\ss}enich, M.~Becker,
  and S.~Apel, ``Preprocessor-based variability in open-source and industrial
  software systems: An empirical study,'' \emph{EMSE}, vol.~21, no.~2, pp.
  449--482, 2016.

\bibitem{Queiroz2015}
R.~Queiroz, L.~Passos, M.~T. Valente, C.~Hunsen, S.~Apel, and K.~Czarnecki,
  ``The shape of feature code: an analysis of twenty c-preprocessor-based
  systems,'' \emph{Software {\&} Systems Modeling}, vol.~16, no.~1, pp. 77--96,
  Jul. 2015.

\bibitem{Hunsen2015}
C.~Hunsen, B.~Zhang, J.~Siegmund, C.~K{\"a}stner, O.~Le{\ss}enich, M.~Becker,
  and S.~Apel, ``Preprocessor-based variability in open-source and industrial
  software systems: An empirical study,'' \emph{EMSE}, vol.~21, no.~2, pp.
  449--482, Apr. 2015.

\bibitem{Spadini2018}
D.~Spadini, M.~Aniche, and A.~Bacchelli, ``{PyDriller}: Python framework for
  mining software repositories,'' in \emph{ESEC/FSE}, 2018.

\bibitem{Zimmermann2007}
T.~Zimmermann, R.~Premraj, and A.~Zeller, ``Predicting defects for eclipse,''
  in \emph{PROMISE}, 2007.

\bibitem{Sliwerski2005}
J.~{\'{S}}liwerski, T.~Zimmermann, and A.~Zeller, ``When do changes induce
  fixes?'' \emph{{ACM} {SIGSOFT} Software Engineering Notes}, vol.~30, no.~4,
  p.~1, Jul. 2005.

\bibitem{Sammut2017}
\emph{Encyclopedia of Machine Learning and Data Mining}.\hskip 1em plus 0.5em
  minus 0.4em\relax Springer {US}, 2017.

\bibitem{Alpaydin2010}
E.~Alpaydin, \emph{{Introduction to Machine Learning}}, second
  edition~ed.\hskip 1em plus 0.5em minus 0.4em\relax Cambridge, Massachusetts:
  The MIT Press, 2010.

\bibitem{domingos2012few}
P.~Domingos, ``A few useful things to know about machine learning,''
  \emph{Communications of the ACM}, vol.~55, no.~10, pp. 78--87, 2012.

\bibitem{hevner2004design}
A.~R. Hevner, S.~T. March, J.~Park, and S.~Ram, ``Design science in information
  systems research,'' \emph{MIS quarterly}, pp. 75--105, 2004.

\bibitem{Ratzinger2008}
J.~Ratzinger, T.~Sigmund, and H.~C. Gall, ``On the relation of refactorings and
  software defect prediction,'' in \emph{MSE}, 2008.

\bibitem{jimenez2019importance}
M.~Jimenez, R.~Rwemalika, M.~Papadakis, F.~Sarro, Y.~Le~Traon, and M.~Harman,
  ``The importance of accounting for real-world labelling when predicting
  software vulnerabilities,'' in \emph{FSE}, 2019.

\bibitem{Chawla2002}
N.~V. Chawla, K.~W. Bowyer, L.~O. Hall, and W.~P. Kegelmeyer, ``{SMOTE}:
  Synthetic minority over-sampling technique,'' \emph{JAIR}, vol.~16, pp.
  321--357, Jun. 2002.

\bibitem{SMK19}
D.~Str{\"{u}}ber, M.~Mukelabai, J.~Kr{\"{u}}ger, S.~Fischer, L.~Linsbauer,
  J.~Martinez, and T.~Berger, ``Facing the truth: benchmarking the techniques
  for the evolution of variant-rich systems,'' in \emph{SPLC}, 2019.

\bibitem{Wen2019}
M.~Wen, R.~Wu, Y.~Liu, Y.~Tian, X.~Xie, S.-C. Cheung, and Z.~Su, ``Exploring
  and exploiting the correlations between bug-inducing and bug-fixing
  commits,'' in \emph{ESEC/FSE}, 2019.

\bibitem{abal201442}
I.~Abal, C.~Brabrand, and A.~Wasowski, ``42 variability bugs in the linux
  kernel: a qualitative analysis,'' in \emph{ASE}, 2014.

\bibitem{leuson2020testconflicts}
L.~D. Silva, P.~Borba, W.~Mahmood, T.~Berger, and J.~Moisakis, ``Detecting
  semantic conflicts via automated behavior change detection,'' in
  \emph{ICSME}, 2020.

\end{thebibliography}
\endgroup

\end{document}